\documentclass[aps,prb,preprint,showkeys,superscriptaddress,groupedaddress]{revtex4-1}  
\usepackage{graphicx}  
\usepackage{dcolumn}   
\usepackage{bm}        
\usepackage{amssymb}
\usepackage{amsmath} 
\usepackage{float} 
\usepackage{placeins}
\begin{document}

%The following information is for internal review, please remove them for submission
%\widetext
%\leftline{Version 01 as of August 2018}
%\leftline{Primary authors: F. Le Roux, D. %D. C. Bradley}
%\leftline{To be submitted to PRB}
%\leftline{Comment to {\tt florian.leroux@physics.ox.ac.uk; donal.bradley@mpls.ox.ac.uk}}

\title{Conformational Control of Exciton-Polariton Physics in Metal - Poly(9,9-dioctylfluorene) - Metal Cavities}
\author{F. Le Roux}
\affiliation{Department of Physics, University of Oxford, Parks Road, Oxford, OX1 3PU, U.K.}
\author{D. D. C. Bradley}
\affiliation{Department of Physics, University of Oxford, Parks Road, Oxford, OX1 3PU, U.K.}
\affiliation{Department of Engineering Science, University of Oxford, Parks Road, Oxford, OX1 3PJ, U.K.}

\date{\today}

\begin{abstract}

Control is exerted over the exciton-polariton physics in metal - Poly(9,9-dioctylfluorene) - metal microcavities via conformational changes to the polymer backbone. Using thin-film samples containing increasing fractions of $\beta$-phase chain segments, a systematic study is reported for the mode characteristics and resulting light emission properties of cavities containing two distinct exciton sub-populations within the same semiconductor. Ultrastrong coupling for disordered glassy-phase excitons is observed from angle-resolved reflectivity measurements, with Rabi splitting energies in excess of 1.05 eV (more than 30$\%$ of the exciton transition energy) for both TE- and TM-polarized light. A splitting of the lower polariton branch is then induced via introduction of $\beta$-phase excitons and increases with their growing fraction. In all cases, the photoluminescence emanates from the lowermost polariton branch, allowing conformational control to be exerted over the emission energy and its angular variation. Dispersion-free cavities with highly saturated blue-violet emission are thus enabled. Experimental results are discussed in terms of the full Hopfield Hamiltonian generalized to the case of two exciton oscillators. The importance of taking account of the molecular characteristics of the semiconductor for an accurate description of its strong coupling behaviour is directly considered, in specific relation to the role of the vibronic structure.
\end{abstract}


\maketitle

\section{Introduction}

The strong coupling regime in semiconductor microcavities,\cite{r1} in which exciton and cavity photon states mix to yield new light-matter excitations (termed microcavity polaritons) has grown into a rich field of research with significant application potential.\cite{r2} Among semiconductor excitons, it has been shown that the Frenkel excitons of organic semiconductors have oscillator strengths sufficient to readily support giant Rabi splittings, $\hbar\Omega_{R}$, of many tens to hundreds of meV\cite{r3,r4,r5,r6,r7,r8,r9,r10,r11} and large enough binding energies ($E_{B}\sim0.5\pm0.25$ eV\cite{r12a,r12b})  that strong coupling can be observed at room temperature.\cite{r2,r3,r4,r5,r6,r7,r8,r9,r10,r11}  

Based on earlier experience with inorganic III-V semiconductors there had been skepticism that organic semiconductor exciton linewidths would allow the observation of strong coupling and for a considerable time following the first demonstration\cite{r3} by Lidzey \textit{et al.} (for porphyrin Soret-band states) the working hypothesis remained that narrow (for organic semiconductors) linewidths would be a necessary pre-requisite. As a consequence, all of the organic systems then studied were selected on that basis, including J-aggregated cyanine dyes, sigma-conjugated polysilanes, porphyrins and napthalene-tetracarboxylic-dianhydride.\cite{r4,r5,r6,r7,r8,r9,r10,r11}  Subsequent studies\cite{r13,r14,r15,r16} on anthracene, ladder poly(p-phenylene)s and phenyl-substituted fluorene trimers showed, instead, that in the presence of such large Rabi splitting the existence of a vibronic progression and significant inhomogeneous broadening was not, in fact, problematic. Indeed, it was in these latter systems that polariton lasing\cite{r14} and Bose-Einstein condensation (BEC)\cite{r15,r16} were first observed and for which the study of condensate physics (e.g. superfluidity of light\cite{r17} and dynamical instability\cite{r18}) has been undertaken. This has opened up a much wider selection of materials for use in strongly-coupled microcavities, many of which have desirable electrical properties, suited to device application.

The specific use of two metal mirrors, that reduce the microcavity mode volume\cite{r19} and provide broader stop bands than for two alternating dielectric layer distributed Bragg reflectors\cite{r1,r13,r14,r15} or a combination of one dielectric and one metal mirror\cite{r3,r4,r5,r6,r7,r8,r9,r10,r11} has additionally led to the observation of Rabi-splitting energies, $\hbar\Omega_{R}$, in excess of 1 eV with operation then occurring in the ultrastrong coupling (USC) regime.\cite{r20,r21,r22} USC physics, for which $\hbar\Omega_{R}$ exceeds a significant fraction ($\geq \sim 20\%$) of the coupled exciton transition energy, yields a number of important differences in spectral features compared to the more commonly observed strong coupling regime\cite{r21,r23a,r23b,r24} and has recently emerged as an interesting path towards novel light-emitting devices\cite{r20,r22,r25a,r25b,r26,r27} and counter intuitive quantum applications, for example the emission of bunched-light from a single quantum emitter.\cite{r28}

Tuning the nature of the photon-exciton coupling for strong coupling systems in order to explore the associated physics and application potential has typically been done by: (i) varying the intra-cavity film thickness, (ii) varying the exciton oscillator strength via dispersion at different concentrations in an inert matrix,\cite{r3,r4,r5} (iii) varying the photon-exciton spatial overlap through normal-to-plane spacer layers,\cite{r5,r6,r7} and (iv) incorporating more than one chemical species.\cite{r5,r29} Other recent approaches include (v) spiropyran-to-merocyanine photoswitching,\cite{r30} (vi) adjusting the mirror-to-mirror distance in an open microcavity\cite{r31,r32} and (vii) reducing dimensionality via the creation of mirror micro-defects.\cite{r33} 

A different approach again is adopted here, namely to use molecular conformation as the vector to control light-matter interactions. Selectively switching the physical structure (conformation) of a fraction of chain segments within a polymer semiconductor film allows us to partition the associated exciton population and systematically tune polariton characteristics and light emission properties. The polymer used to do this, namely poly(9,9-dioctylfluorene) (PFO), has been employed to good effect in the fabrication of a wide range of devices, including Light-Emitting Diodes (LEDs),\cite{r34} transistors\cite{r35} and optically-pumped lasers.\cite{r36a}\cite{r36b}\cite{r36c} Formation of so-called $\beta$-phase, linearly-extended, planar chain segments in PFO leads to red-shifted, characteristically well-resolved, vibronically-structured absorption and PL emission bands, with a small Stokes’ shift.\cite{r37a,r37b,r37c,r38,r39a,r39b,r40} The associated refractive index and optical gain changes are similarly notable.\cite{r41,r42a,r42b,r43,r44,r45} This conformation change can be readily generated in otherwise glassy-phase films by thermally cycling or solvent vapour annealing,\cite{r37a,r37b,r37c,r38,r41,r44} slow drying,\cite{r39a,r39b} and swelling with solvent or solvent/non-solvent mixtures.\cite{r40,r45} Another approach, as used below, has been to add controlled amounts of 1,8-diiodooctane\cite{r46} or paraffin oil\cite{r47} to the polymer solution prior to spin-coating, allowing a systematic variation in $\beta$-phase fraction without the need for post-deposition treatment. This conformational tunability allows detailed studies of photophysics, including energy migration processes\cite{r48} and dichroism.\cite{r38} In addition, it provides a metamaterials structuring approach to patterning the refractive index in order to generate photonic structures.\cite{r44,r45}

In the following, the different coupling regimes, consequent mode characteristics and resulting light emission properties are explored for glassy- and $\beta$-phase exciton populations (labelled $X_{g}$ and $X_{\beta}$) present within Al-PFO-Al microcavities. 
Angle-resolved reflectivity and steady-state photoluminescence (PL) are used to demonstrate USC for 100$\%$ $X_{g}$ with $\hbar\Omega_{R_{TE,TM}}>1.05$ eV (i.e. $>$ 30$\%$ of the $X_{g}$ optical transition peak energy = 3.25 eV). The addition of a $X_{\beta}$ population to which the cavity strongly couples then leads to a further splitting of the lower polariton (LP) branch that increases with $X_{\beta}$ fraction (0 to 15.8$\%$ studied here), allowing systematic control over the ensuing exciton-polariton physics. As expected, PL emanates from the new lowermost polariton branch, enabling conformation tuning of the emission energy and its angular dispersion. This approach offers the prospect of viewing-angle-independent, narrow-emission-linewidth, microcavity light-emitting diodes benefiting from both USC and conformationally-controlled strong coupling. 

The experimental results are analysed using the Agranovich-Hopfield model,\cite{r23a,r23b,r24} as adopted in previous molecular microcavity USC studies\cite{r20,r21,r22,r24,r25a,r25b,r26} but generalized to the case of two distinct exciton populations. Analysis of the Rabi splitting energies in relation to exciton oscillator strengths underlines the importance of accounting for vibronic structure. The discussion focuses on recent observations\cite{r49} and theoretical approaches based on the use of the Holstein-Tavis-Cummings (HTC) model,\cite{r50a,r50b,r50c,r51} finally highlighting the need for further theoretical efforts to help realize a new generation of quantum devices based on conformationally-controlled exciton-polariton physics.

\section{PFO Thin Film Optical Properties}

\subsection{Absorption and Photoluminescence Spectra}

PFO thin film samples (PL0388 from 1-Material Inc\cite{r52} with weight average molecular weight, Mw = 55,000 and polydispersity index, PDI = 2.5) were deposited on pre-cleaned fused silica substrates by spin coating (2300 rpm for one minute). 1-8 diiodooctane was first added to toluene in order to fabricate 8 precursory mixtures (with volume concentrations ranging from 0 to 35 $\mu$L per mL). 100 $\mu$L of precursory mixture was then mixed into 1.9 mL of the baseline 13.5 mg per mL PFO in toluene solution, resulting in increasing amounts (0 to 0.175 vol$\%$) of 1-8 diiodooctane. Both the solutions and the substrates were then pre-heated at 80$^{\circ}$C for 5 minutes immediately prior to deposition. The thicknesses of the resulting films were measured, using a Dektak profilometer, to be 75-85 $\pm$ 2 nm. Samples were prepared with eight different $\beta$-phase fractions (increasing together with increasing vol$\%$ of 1-8 diiodooctane), namely 0, 1.1 $\pm$ 0.5, 1.5 $\pm$ 0.5, 3.7 $\pm$ 0.5, 10.0 $\pm$ 0.5, 12.3 $\pm$ 0.5, 13.9 $\pm$ 0.5 and 15.8 $\pm$ 0.5$\%$, determined from the relative strength of the duly weighted (to correct for the known oscillator strength increase), integrated $X_{\beta}$ absorption contribution.\cite{r53} These fractions are used as reference labels for both films and corresponding microcavities in the rest of the paper.

\begin{figure}[h]
\includegraphics[scale=0.34]{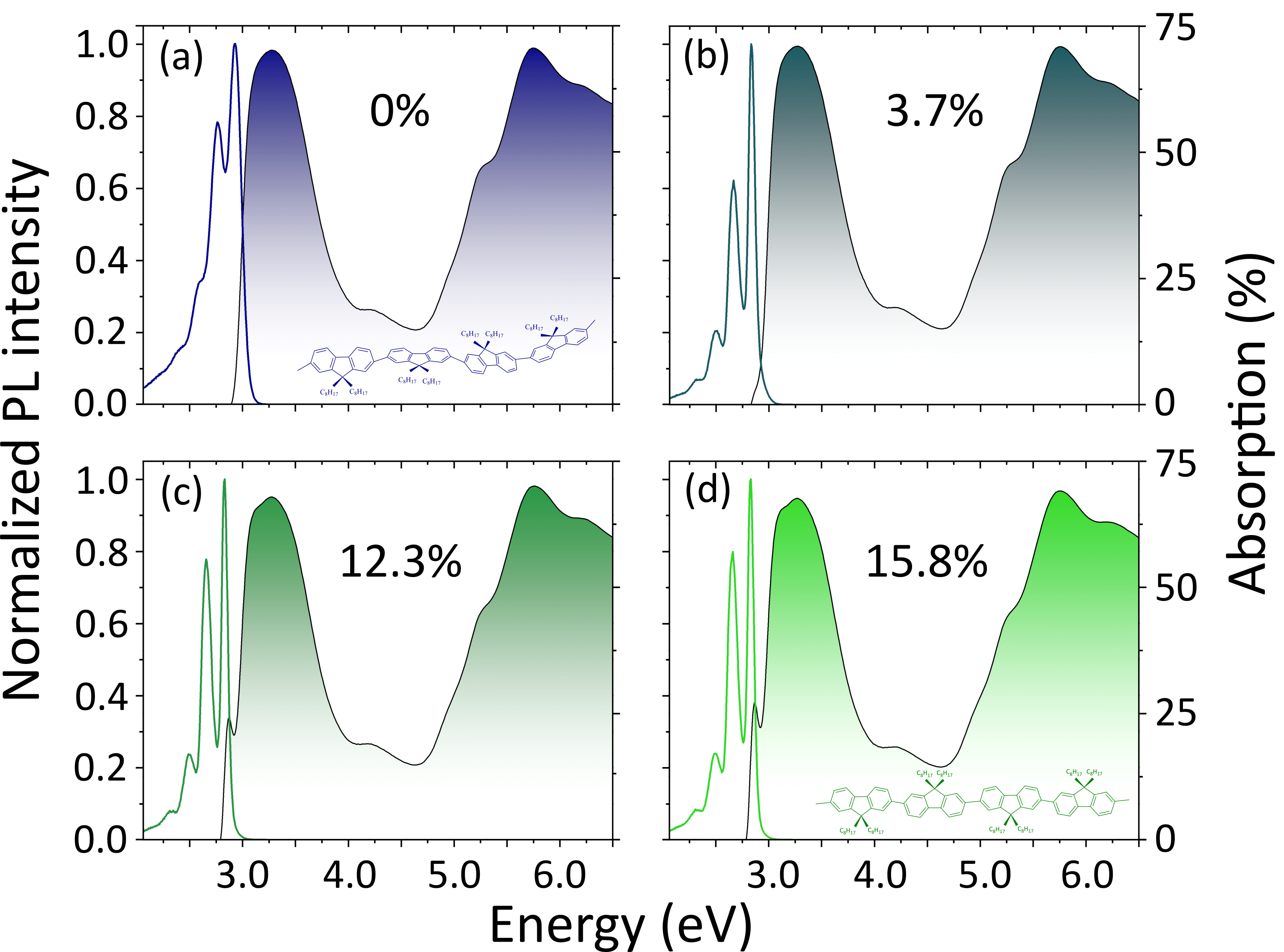}
\caption{\label{fig:Figure1} Absorption and PL spectra for spin-coated PFO films containing: (a) 0 (glassy-phase), (b) 3.7, (c) 12.3, and (d) 15.8$\%$ $\beta$-phase chain segments. Note the emergence of a shoulder (b) and then a resolved peak ((c) and (d)) on the red edge of the absorption with increasing $\beta$-phase fraction. A schematic glassy-phase chain segment (one conformer of many) is shown in panel (a) and the $\beta$-phase chain extended structure in panel (d).}
\end{figure}

Thin film absorption spectra were measured using a Perkin-Elmer Lambda 1050 UV/Vis spectrophotometer and were corrected for specular reflectance. FIG.~\ref{fig:Figure1} shows data for four samples, namely those with 0, 3.7 $\pm$ 0.5, 12.3 $\pm$ 0.5 and 15.8 $\pm$ 0.5$\%$ $\beta$-phase fractions, whilst the corresponding spectra for samples with the other four fractions are presented in Supplemental Material (FIG. S1).\cite{SI} In FIG.~\ref{fig:Figure1} (a) the 0$\%$ (glassy-phase) absorption spectrum can be described, for simplicity, as comprising two dominant contributions, namely a lower energy delocalized $X_{g}$ $S_0-S_1$ exciton band (peaked at $\sim$ 3.25 eV) and higher energy absorption (shoulder at $\sim$ 5.28 eV, peak at $\sim$ 5.71 eV and higher energy shoulder at $\sim$ 6.25 eV) associated with ring-localized fluorene states;\cite{r37c} the additional absorption at ~ 4.25 eV is much weaker. A schematic glassy-phase chain segment is shown at the bottom of panel (a), with deviations from the mean fluorene-fluorene single bond torsion angle ($\sim$ 135$^{\circ}$) generating multiple conformers that cause inhomogeneous broadening of the $X_{g}$ $S_0-S_1$ absorption.\cite{r39b} The expected vibronic structure is not then resolved, with the full width at half maximum (FWHM) linewidth of the transition $\approx$ 0.6 eV.

Formation of $\beta$-phase chain segments (c.f. schematic structure at the bottom of panel (d)) leads to the appearance of a characteristic (0-0) vibronic absorption peak at 2.87 $\pm$ 0.01 eV, superimposed on the red-edge of the $X_{g}$ band.\cite{r37a,r37b,r37c} This absorption grows in proportion to the film $\beta$-phase fraction (c.f. panels (c) and (d)) and is accompanied by additional spectral changes associated with the higher vibronics of the $X_{\beta}$ $S_0-S_1$ optical transition, that lie in the vicinity of the $X_{g}$ absorption peak.\cite{r37a,r37b,r37c}  

PL spectra were measured using a Horiba Fluorolog spectrofluorometer with excitation wavelength set to 3.31 eV (375 nm). For both glassy- and $\beta$-phase samples, the thin-film PL spectra reveal relatively well-resolved vibronic structures; absorption probes the whole ensemble of chain segment conformations whilst PL occurs only from a self-selected subset of lower energy sites following exciton migration within the ensemble. In the glassy case (c.f. FIG.~\ref{fig:Figure1} (a)), the $X_{g}$ $S_{1}-S_{0}$ (0-0), (0-1) and (0-2) PL vibronic peaks appear at 2.93 eV (423 nm), 2.77 eV (448 nm) and 2.58 eV (481 nm) whereas following $\beta$-phase chain segment formation the PL vibronic peaks red-shift to 2.83 eV (438 nm), 2.66 eV (467 nm) and 2.49 eV (497 nm), also becoming better resolved (c.f. FIG.~\ref{fig:Figure1} (b), (c), (d)).  $\beta$-phase chain segments have a reduced optical gap on account of their enhanced conjugation and their well-defined nature leads to the observed reduction in inhomogeneous broadening.\cite{r37c,r48} Efficient and rapid ($\sim$ ps) exciton transfer from glassy- to $\beta$-phase chain segments then leads to $X_{\beta}$ emission dominating the PL spectrum even at relatively low $\beta$-phase fractions;\cite{r40,r48} $\beta$-phase segments act as a self-dopant to which glassy-phase excitations are funneled.\cite{r54}

\subsection{Optical Constants}

The optical constants for all eight $\beta$-phase fraction films (0, 1.1, 1.5, 3.7, 10.0, 12.3, 13.9 and 15.8 $\%$), were determined using Variable Angle Spectroscopic Ellipsometry (VASE) with a J.A. Woollam RC2 ellipsometer. For each sample, three reflection-geometry measurements were performed with light incident at 45$^{\circ}$, 50$^{\circ}$, and 55$^{\circ}$ (angles of incidence are quoted relative to the plane normal), together with a normal incidence (0$^{\circ}$) transmittance measurement. Recognizing that PFO chains tend to lie preferentially within the plane of the film\cite{r55a,r55b,r56} a uniaxial anisotropic model was used to fit these data and extract the in-plane (ordinary), ($n_{xy}$, $k_{xy}$), and out-of-plane (extra-ordinary), ($n_{z}$, $k_{z}$), spectral components of the complex refractive index $\tilde{n} = n + ik$. Interferometric enhancement of the measurement sensitivity allows a more accurate determination of the refractive index anisotropy\cite{r56} but that is not a key concern for the current work. Previous studies have reported ellipsometry-derived values for ($n_{xy}$, $k_{xy}$) and ($n_{z}$, $k_{z}$) for glassy PFO\cite{r55a,r55b,r57} and have modeled ($n_{xy}$, $k_{xy}$) for different $\beta$-phase fractions, using a semi-empirical approach based on the separability of $X_{g}$ and $X_{\beta}$ contributions.\cite{r44,r44,r45} Our experiments agree with and extend those earlier results by providing experimental data for a systematic and wide ranging variation in $\beta$-phase fraction. 

Thin film optical constants determined for four $\beta$-phase fractions are shown in FIG.~\ref{fig:Figure2}, with data for the other four films presented in Supplemental Material (FIG. S2).\cite{SI} As expected, emergence of the $X_{\beta}$ (0-0) vibronic peak in k leads to a correspondingly significant enhancement in n across the PL spectral range.\cite{r41,r44} As a direct consequence, spatial patterning of the $\beta$-phase fraction, for example using dip pen nanolithography, then allows the fabrication of photonic structures such as laser gratings.\cite{r45}
\newpage
\begin{figure}[H]
\includegraphics[scale=0.34]{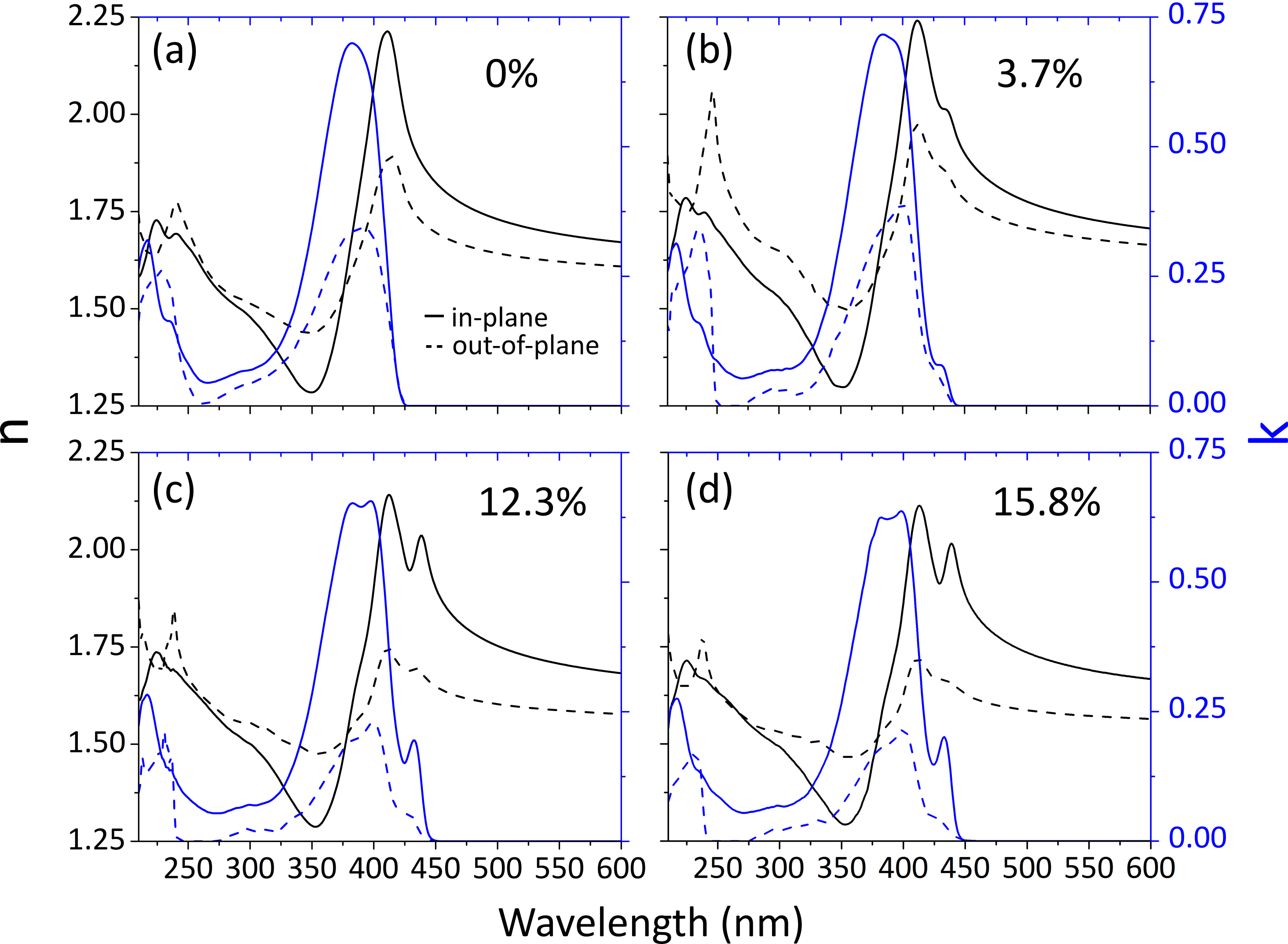}
\caption{\label{fig:Figure2} Optical constants for spin-coated PFO thin films containing: (a) 0 $\%$ (glassy-phase), (b) 3.7, (c) 12.3, and (d) 15.8$\%$ $\beta$-phase chain segments. The black and blue lines are, respectively, the real (n) and imaginary (k) parts of the complex refractive index, $\tilde{n} = n + ik$. Solid lines give in-plane ($n_{xy}$, $k_{xy}$) and dashed lines out-of-plane ($n_{z}$, $k_{z}$) values.
}
\end{figure}
\newpage
\section{Exciton-Polariton Physics: Measurement and Simulation}

\subsection{Microcavity Fabrication}

Transfer Matrix Reflectivity (TMR) calculations, based on the optical constants presented in FIG.~\ref{fig:Figure2}, were used to design the Al-PFO-Al microcavity structures and to interpret the observed angle-resolved reflectivity and PL results. Microcavities consisting of films of PFO sandwiched between 100 nm bottom and 23 nm top Al mirrors were fabricated for all eight $\beta$-phase fractions. The mirrors were thermally evaporated at a base pressure of $\sim 10^{-7}$ mbar on cleaned fused silica substrates (bottom mirror) and on top of the PFO film (top mirror); the latter was spin-coated onto the bottom mirror from toluene solutions containing varying concentrations of 1,8-diiodooctane additive, as described in Section II.A. The resulting polymer film thicknesses were determined by fitting TMR calculations to the reflectivity spectra, yielding 82 to 92 $\pm$ 2 nm.

\subsection{Angle-Resolved Reflectivity Spectra for Varying $\beta$-phase Fractions}

Reflectivity spectra (across the energy range 2.5 to 5 eV) were measured for incidence angles between 45 and 75$^{\circ}$ using a Woollam RC2 ellipsometer and plotted as R intensity maps on an energy vs angle plane. Both TE- and TM-polarizations were measured since they show significantly different behaviour.\cite{r58} The TE-polarized results for four (0$\%$, 3.7$\%$, 12.3$\%$, and 15.8$\%$) of the eight microcavities are shown in FIG.~\ref{fig:Figure3}, with the other four datasets in Supplemental Material (FIG. S3).\cite{SI} For each map, the reflectivity minima (polariton energy levels) obtained via TMR calculations (white dashed-lines) using the optical constants from FIG.~\ref{fig:Figure2} are overlaid on the experimental dispersions (presented as colour-coded R intensities). There are two resolved levels (Lower Polariton (LP) and Upper Polariton (UP)) for $\beta$-phase fractions $\leqslant 1.5 \%$, with three levels (LP, Middle Polariton (MP), UP) evident for $\beta$-phase fractions $\geqslant 3.7 \%$; the use of these labels is discussed in Section IV below. For each panel, the inset shows the spectral dispersion of the lowest polariton branches (LP only for (a), LP and MP for (b), (c) $\&$ (d)) plotted between 2.80 and 2.98 eV for incidence angles, from bottom to top, of 45, 60 and 75$^{\circ}$.  In addition, to the right of each panel is the corresponding, colour-rendered, in-plane extinction coefficient spectrum. 
\begin{figure}[H]
\centering
\includegraphics[scale=0.5]{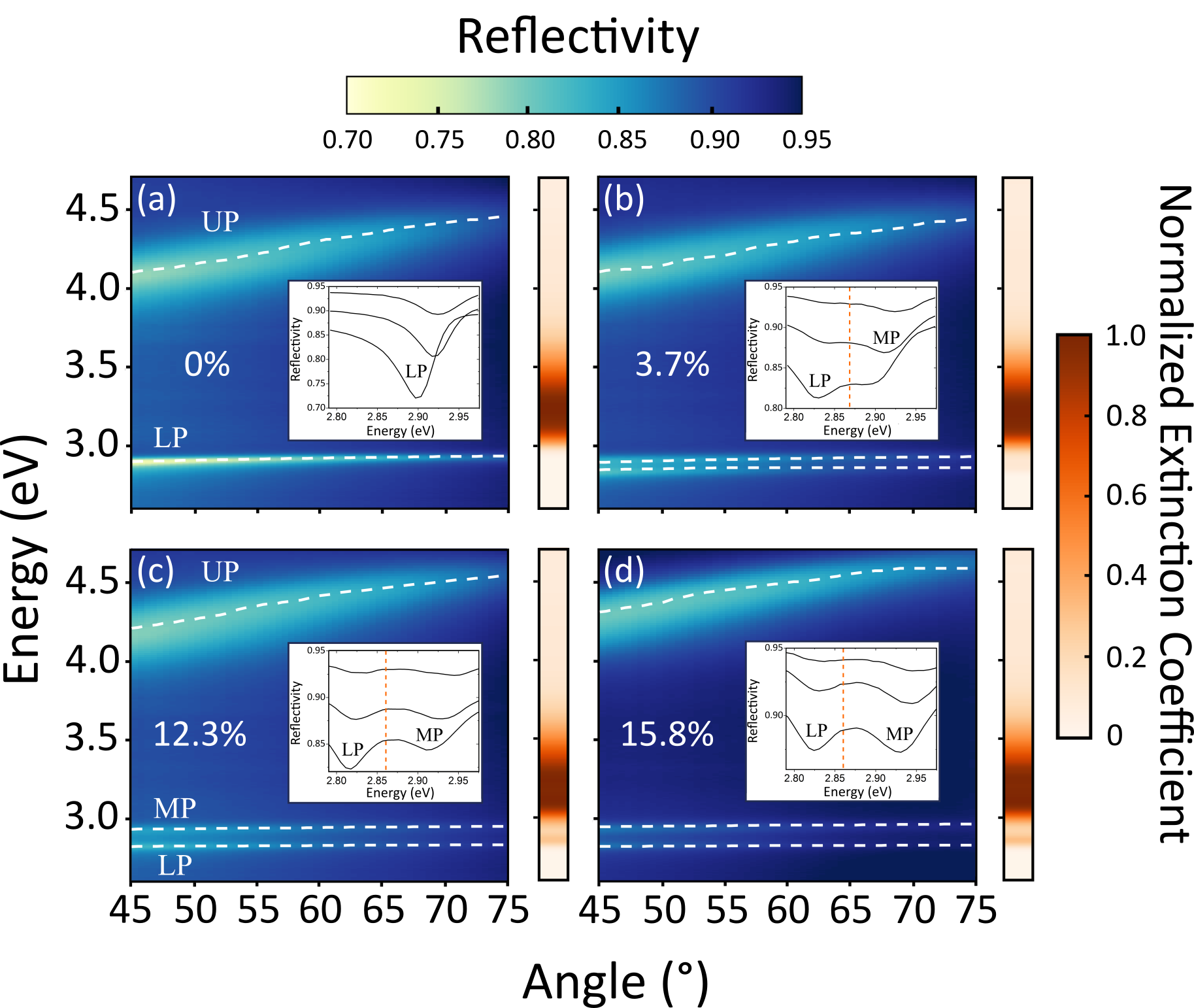}
\caption{\label{fig:Figure3} Experimental, angle-resolved, TE-polarized reflectivity maps for microcavities containing (a) 0, (b) 3.7, (c) 12.3, and (d) 15.8$\%$ $\beta$-phase chain segments. The inset to each panel shows the spectral dispersion of the polariton branches between 2.80 and 2.98 eV for incidence angles (bottom to top) 45, 60 and 75$^{\circ}$. Note the splitting of the LP into two separate branches (LP and MP) for $\geqslant$ 3.7$\%$ $\beta$-phase fraction (panels (b), (c) $\&$ (d)), spaced either side of the $X_{\beta}$ (0-0) vibronic peak at 2.87 eV (3.7$\%$) and 2.86 eV (12.3$\%$, 15.8$\%$) (orange dashed vertical line). The corresponding normalized, in-plane, bare-film, colour-rendered extinction coefficient spectra taken from FIG.~\ref{fig:Figure2} are shown to the right of each map. Also shown, as overlaid white dashed lines, are the reflectivity minima dispersion curves obtained from TMR calculations. 
}
\end{figure}
From FIG.~\ref{fig:Figure3} (a) the energy splitting between the $X_{g}$-derived LP and UP branches clearly exceeds 1 eV (analytical derivation of the Rabi energy is perfomed in Section IV). The LP also has a much narrower FWHM linewidth (0.05 $\pm$ 0.005 eV at 45$^{\circ}$) than does the UP (0.31 $\pm$ 0.05 eV), generally a signature of motional narrowing effects within the inhomogeneously broadened $S_0-S_1$ transition as the photon mode energy gets closer to the centre of the excitonic distribution (c.f. extinction coefficient spectrum to right of panel).\cite{r59} Quantification of the associated coupling parameters is performed in Section IV.

The R-map in FIG.~\ref{fig:Figure3} (b), for the 3.7 $\pm$ 0.5$\%$ $\beta$-phase fraction film, shows the appearance of an additional polariton branch. The resulting MP branch lies relatively close to the LP branch of the 0$\%$ cavity with a new LP branch emerging at lower energy. Analogous behaviour is seen in FIG.~\ref{fig:Figure3} (c) and (d) with an increasing LP to MP energy separation as the $\beta$-phase fraction grows; detailed analysis of this behaviour is presented in Section IV.  The panel insets show this splitting in greater detail, presenting energy cross-sections at fixed angles.

For the 12.3$\%$ $\beta$-phase cavity (FIG.~\ref{fig:Figure3} (c) inset), the LP FWHM at 45$^{\circ}$ was fitted to be 0.053 $\pm$ 0.01 eV, with the error largely arising from the proximity of the MP. The LP R-dip at this angle is significantly deeper than that for the MP but with increasing angle, equivalent to increasing the photon mode energy, the relative intensity of the R-dips reverses and the MP becomes the deeper one. This is entirely as expected for the anti-crossing behaviour observable in strong coupling.

\subsection{Angle-Resolved Photoluminescence Emission}

\begin{figure}[H]
\begin{flushleft}
\includegraphics[scale=0.5]{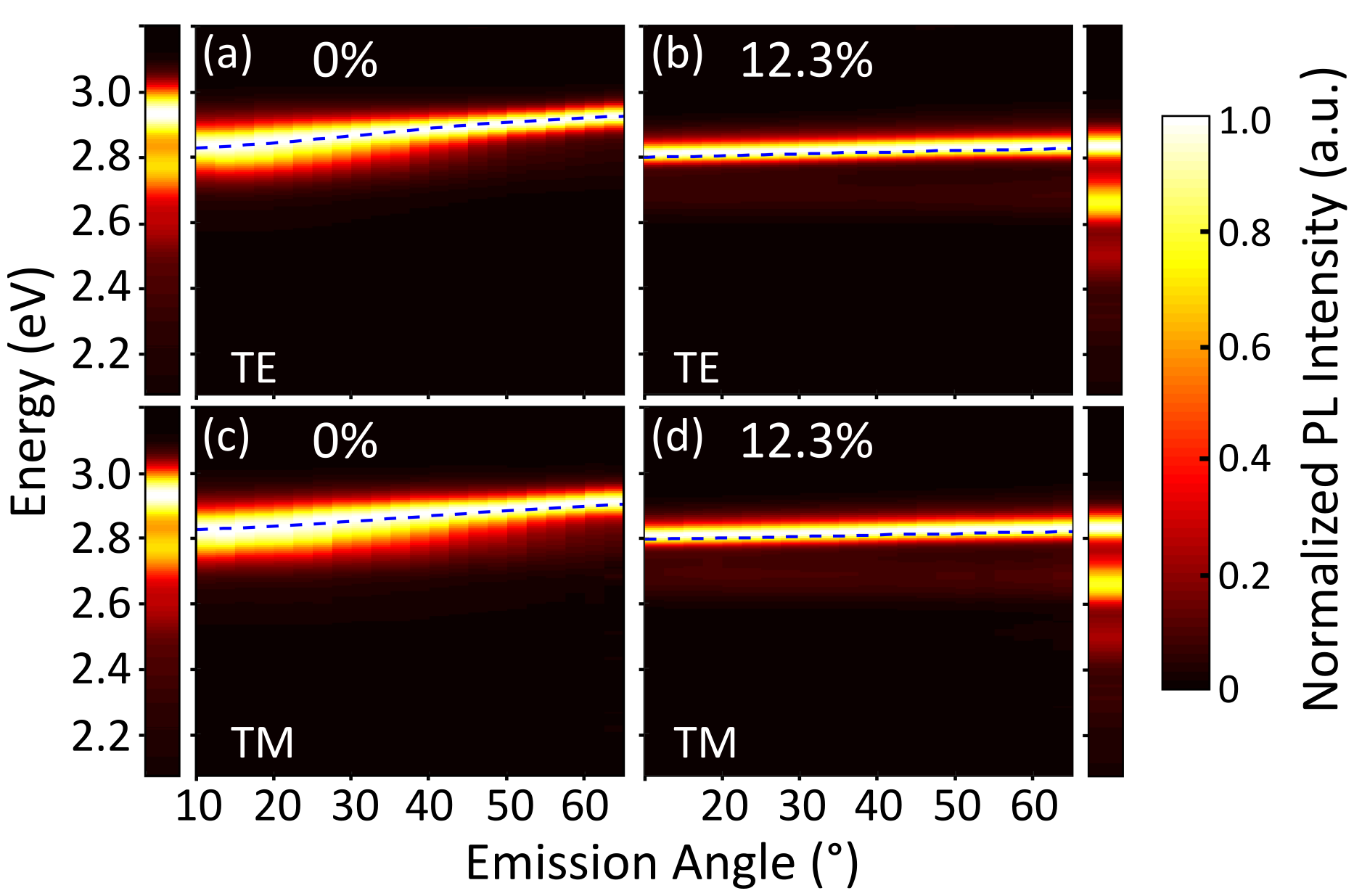}
\end{flushleft}
\caption{\label{fig:Figure4} Angle-resolved PL spectral intensity maps for microcavities containing 0$\%$ ((a) $\&$ (c)) and 12.3$\%$ ((b) $\&$ (d)) $\beta$-phase chain segments, with TE ((a) $\&$ (b)) and TM ((c) $\&$ (d)) polarized spectra plotted separately. The colour-rendered spectra to the left of (a) $\&$ (c) and right of (b) $\&$ (d) present the PL emission obtained for the corresponding bare films (FIG.~\ref{fig:Figure1} (a) $\&$ (c)). The overlaid blue dashed line is the LP angular dispersion obtained from TMR calculations.}
\end{figure}

Angle-resolved PL emission was measured using a Horiba Fluorolog spectrofluorometer, with excitation energy set to 3.31 eV (375 nm). Spectra were recorded for emission angles between 10 and 65$^{\circ}$ at 2.5$^{\circ}$ steps and a polarizer was used to separate TE and TM polarizations. The excitation energy was chosen to optically pump the intense absorption line arising predominantly from glassy-phase $X_{g}$ $S_{0}-S_{1}$ optical transitions; the pump wavelength selection will be further discussed in Section V.B. below.  FIG.~\ref{fig:Figure4} shows normalized PL intensity maps plotted on energy vs angle planes for both TE- ((a) and (b)) and TM-polarized ((c) and (d)) emission from 0$\%$ ((a) and (c)) and 12.3$\%$ ((b) and (d)) $\beta$-phase microcavities. The dashed blue lines are TMR calculations of the LP dispersion as per FIG.~\ref{fig:Figure3}. The colour-rendered free-space unpolarised PL intensity spectra for bare films are plotted to the left and right of their respective PL maps; the panels to the left of (a) and (c) corresponding to 0$\%$ and the panels to the right of (b) and (d) to 12.3$\%$ $\beta$-phase fraction. In all cases the microcavity emission is dominated by a single peak originating from the LP, both in the presence and absence of $\beta$-phase chain segments. 

At 45$^{\circ}$, for 0$\%$ $\beta$-phase, the TE- and TM-polarized emission FWHM linewidths are both $\Delta E_{FWHM} = 91\pm5$ meV. This is substantially ($>$ six-fold) narrower than for PFO bare film $X_{g}$ PL emission - where $\Delta E_{FWHM}\approx600 meV$, determined by the inhomogeneously broadened vibronic progression - and narrower even than the deconvolved (0-0) vibronic peak for which $\Delta E_{FWHM}\approx135$ meV. It is shown further in Section IV that for TE polarization this linewidth is also narrower than the photon mode linewidth (145 meV) of an equivalent cavity with an effective refractive index $n_{eff}$ = 1.6, consistent with results obtained in similar studies elsewhere.\cite{r20,r21,r22} At the same angle, for 12.3$\%$ $\beta$-phase fraction cavities, both TE- and TM-polarized $\Delta E_{FWHM} = 50 \pm 3$ meV, comparable to the width of the LP (53 $\pm$ 10 meV) obtained from reflectivity, but narrower than the resolved (0-0) vibronic of the $X_{\beta}$ PL emission from bare films ($\Delta E_{FWHM} = 76 \pm 2$ meV). It is also almost three times narrower than the photon mode width (again 145 meV). 

For the 0$\%$ cavity, the LP emission gradually blue-shifts from 2.82 eV to 2.92 eV, for both polarizations, as the angle increases from 10 to 65$^{\circ}$. This relatively weak dispersion is directly comparable to that seen in other USC studies.\cite{r20,r21,r22} Introduction of $X_{\beta}$ strong coupling further diminishes the LP dispersion and for the 12.3$\%$ cavity the LP is essentially dispersionless, with PL emission then becoming angle insensitive. As previously noted,\cite{r8} this is of direct interest for the fabrication of high spectral-purity LEDs that would not suffer from the significant angular blue-shift that is typical for weakly-coupled microcavity structures.\cite{r60a,r60b} Further details are given in Supplemental Material FIG. S5 and accompanying text.\cite{SI} In addition, although as a consequence of LP lineshape asymmetry there is weak (accounting for $<10\%$ of the total) lower energy emission from the 12.3$\%$ cavity, this would not prevent the achievement of an exceptionally-saturated, deep-blue/violet LED emission. The measured LP PL spectrum corresponds to colour coordinate CIE (x,y) = (0.161,0.018), with dominant wavelength 447 nm and 99$\%$ saturation (FIG. S5). 

\section{Theoretical Analysis}

The large ($>$ 1eV) LP to UP splitting of the polariton energy levels relative to the centre of the $X_{g}$ optical transition (3.25 eV) firmly suggests ultrastrong coupling, a hypothesis that is confirmed \textit{a posteriori} by determining the corresponding Rabi energy through comparison to a suitable model, such as that developed by Agranovich and Hopfield.\cite{r23a,r23b} In this study, the basic model is extended to the case of two distinct exciton populations ($X_{g}$ and $X_{\beta}$) simultaneously coupled to the cavity. Within the model, these populations are treated as spectrally localized oscillators (1 and 2) whereas in reality their oscillator strengths are rather more spread in energy, across resolved ($X_{\beta}$) and un-resolved ($X_{g}$) vibronic progressions that spectrally overlap. The nature of the cavity photon - exciton coupling is, therefore, subject to further consideration in Section V.B. but for now the theoretical analysis proceeds along the simpler (spectrally localized oscillator) path in order to more readily explore the essential features of the governing physics.
The full Hamiltonian is then the sum of three terms: 
\begin{equation} \label{eq:1}
H= H_{0}+H_{res}+ H_{anti} 	
\end{equation}								
with 
\begin{equation} \label{eq:2}
H_{0} = \hbar\sum_{q}(\omega_{cav_q}(a^{\dagger}_qa_q+\frac{1}{2})+\sum_{j}\omega_{j}(b^{\dagger}_{j,q}b_{j,q}+\frac{1}{2})),
\end{equation}
\begin{equation} \label{eq:3}
H_{res} = \sum_{q}(D_{q}(a_qa_{-q}+a^{\dagger}_qa_{-q}                                                                                                                                                                                                                                            )+i\hbar\sum_{j}\frac{\Omega_{j,q}}{2}(a_{q}b_{j,-q}-a^{\dagger}_{q}b^{\dagger}_{j,-q})),
\end{equation}
\begin{equation} \label{eq:4}
H_{anti} = \sum_{q}(D_{q}(a^{\dagger}_{q}a_{q}+a_{q}a^{\dagger}_{q})                                                                                                                                                                                                                                         )+i\hbar\sum_{j}\frac{\Omega_{j,q}}{2}(a^{\dagger}_{q}b_{j,q}-a_{q}b^{\dagger}_{j,q})).
\end{equation}				              	  
In Eq.~(\ref{eq:2}),~(\ref{eq:3}),~(\ref{eq:4}), q is the in-plane wave vector, $a_q$  and $a^{\dagger}_{q}$ respectively annihilate and create a photon at frequency $\omega_{cav_q}$, $b_j$  and $b^{\dagger}_j$ respectively annihilate and create a j-exciton at frequency $\omega_j$ with $j$ $\epsilon$ $\left \{1,2\right\}$,  $\Omega_{j,q}$ is the associated Rabi frequency, and for a given angle $\theta$: $\Omega_{j,q}= \Omega_{j}(\theta) = \Omega_{0j}\sqrt{\frac{\omega_j}{\omega_{cav}(\theta)}}$ where $\Omega_{0j}$ is the Rabi frequency on resonance for the j-excitons. It was shown that in metal-organic semiconductor-metal cavities $\omega_{cav}(\theta)$ can be approximated by \cite{r20} 
\begin{equation}\label{eq:5}
\omega_{cav_{(TE,TM),q}} = \omega_{cav_{(TE,TM)}}(\theta)= 
\omega_{cav}(0)(1-\frac{sin^{2}(\theta)}{n^{2}_{eff_{TE,TM}}})^{-\frac{1}{2}}
\end{equation}
where $n_{eff_{TE,TM}}$ is polarization dependent. Finally, $D_q = \sum _{j} \frac{\Omega_{j,q}^2}{4\omega_{j}}$ is the contribution of the squared magnetic vector potential.

In order to diagonalize \textit{H}, the polariton annihilation operators $p_{i,q} = w_{i,q}a_{q}+\sum_{j}x_{i,j,q}b_{j,q}+y_{i,q}a^{\dagger}_{-q}+\sum_{j}z_{i,j,q}b^{\dagger}_{j,-q}$ for $i\epsilon\left \{LP,MP,UP\right\}$ are introduced, where w, x, y and z label, respectively, the photon, exciton, anomalous photon and anomalous exciton Hopfield coefficients. After adding the constant terms into the ground state energy $E_G$, \textit{H} can be diagonalized in the form:
\begin{equation}\label{eq:6}
H = E_{G} + \sum_{i\epsilon\left \{LP,MP,UP\right\}}\sum_{q}\hbar\omega_{i,q}p^{\dagger}_{i,q}p_{i,q},
\end{equation}						
and is obtained provided that:
\begin{equation}\label{eq:7}
\overrightarrow{v}_{i,q}=(w_{i,q},x_{i,1,q},x_{i,2,q},y_{i,q},z_{i,1,q},z_{i,2,q})^{T}
\end{equation}
is a solution to the eigenvalue problem:
\begin{equation}\label{eq:8}
M_q \overrightarrow{v}_{i,q}= \omega_{i,q} \overrightarrow{v}_{i,q},
\end{equation}									
where $M_q$ reads
\begin{equation}\label{eq:9}
M_{q}=
\begin{bmatrix}
\omega_{cav,q}+2D_{q} & -i\frac{\Omega_{1,q}}{2} & -i\frac{\Omega_{2,q}}{2} & -2D_{q} & -i\frac{\Omega_{1,q}}{2} & -i\frac{\Omega_{2,q}}{2}\\ 
i\frac{\Omega_{1,q}}{2} & \omega_{1} & 0 & -i\frac{\Omega_{1,q}}{2} & 0 & 0\\
i\frac{\Omega_{2,q}}{2} & 0 & \omega_{2} & -i\frac{\Omega_{2,q}}{2} & 0 & 0\\ 
2D_{q} & -i\frac{\Omega_{1,q}}{2} & -i\frac{\Omega_{2,q}}{2} & -\omega_{cav,q} - 2D_{q} & -i\frac{\Omega_{1,q}}{2} & -i\frac{\Omega_{2,q}}{2}\\
-i\frac{\Omega_{1,q}}{2} & 0 & 0 & i\frac{\Omega_{1,q}}{2} & -\omega_{1} & 0\\ 
-i\frac{\Omega_{2,q}}{2} & 0 & 0 & i\frac{\Omega_{2,q}}{2} & 0 & -\omega_{2}
\end{bmatrix}   ,
\end{equation}           
with $\omega_{i,q}\epsilon\left \{\omega_{LP,q},\omega_{MP,q},\omega_{UP,q},-\omega_{LP,q},-\omega_{MP,q},-\omega_{UP,q}\right\}$. In the case of a single exciton oscillator, $M_q$ reduces to the usual 4 x 4 Hopfield-like USC matrix.\cite{r20,r24,r26}
The Bose commutation rule $([p_{i,q},p^{\dagger}_{i,q}]=\delta_{i,i'}\delta_{q,q'})$ fully defines the problem by further imposing:
\begin{equation}\label{eq:10}
\left | w_{i,q} \right |^{2}+\sum_{j}\left | x_{i,j,q} \right |^{2} - \left | y_{i,q} \right |^{2} - \sum_{j}\left | z_{i,j,q} \right |^{2} = 1.	
\end{equation}	
The eigenvalues of $M_q$ were fitted to the experimental results for each cavity, for both TE- and TM-polarization, using the PL maxima in the 10 - 45$^{\circ}$ range from FIG.~\ref{fig:Figure4} together with the R-minima in the 45 - 75$^{\circ}$ range from FIG.~\ref{fig:Figure3}. The 4 x 4 Hopfield-like USC matrix was used for the 0$\%$ microcavity, while the full matrix $M_q$ was used for $\beta$-phase fractions $\geqslant3.7\%$. The 1.1 and 1.5 $\%$ $\beta$-phase microcavities were left out of the initial analysis as the $\beta$-phase content is too small/insufficiently spectrally separated to induce a splitting of the LP branch. They are briefly discussed in Section V in light of the results for the other microcavities.

In order to minimize the number of fitting parameters and obtain meaningful results, only $\omega_{cav_{TE,TM}}(0)$, $n_{eff_{TE,TM}}$  $\&$  $\Omega_{01_{TE,TM}}$ were allowed to vary in fitting the 0$\%$ microcavity. Similarly, only $\omega_{cav_{TE,TM}}(0)$, $n_{eff_{TE,TM}}$, $\Omega_{01_{TE,TM}}$ $\&$  $\Omega_{02_{TE,TM}}$ were allowed to vary in fitting the other samples. The value of $\hbar\omega_{1}$ was set to be at the energy that corresponds to the mid-point of the integral oscillator strength for the $X_{g}$ $S_{0}-S_{1}$ optical transition using $\int_{2.8}^{\hbar\omega_{1}}\epsilon(\omega)d\omega=\frac{1}{2}\int_{2.8}^{3.7}\epsilon(\omega)d\omega$, where $\epsilon(\omega)$ is the extinction coefficient for $X_{g}$ in the 2.8 to 3.7 eV energy range; this yields $\hbar\omega_{1} = 3.25$ eV. Conversely $\hbar\omega_{2}$ was set to the $S_{0}-S_{1}$ (0-0) vibronic peak energy of the $X_{\beta}$ in-plane extinction coefficient, namely 2.87 eV for 3.7 $\&$ 10$\%$ cavities and, reflecting a slight red-shift with increasing $\beta$-phase content, 2.86 eV for 12.3, 13.9 $\&$ 15.8$\%$ cavities.\cite{r37a,r37b,r37c} This choice for $\hbar\omega_{2}$ is discussed further in Section V.B. and arises from the dominant contribution that the $\beta$-phase (0-0) vibronic transition appears to make when the 0$\%$ LP splits into MP and LP. It allows good fits to the data, whereas the equivalent integral to that used for $\hbar\omega_{1}$ gives inconsistent results. 

\begin{figure}[H]
\begin{flushleft}
\includegraphics[scale=0.5]{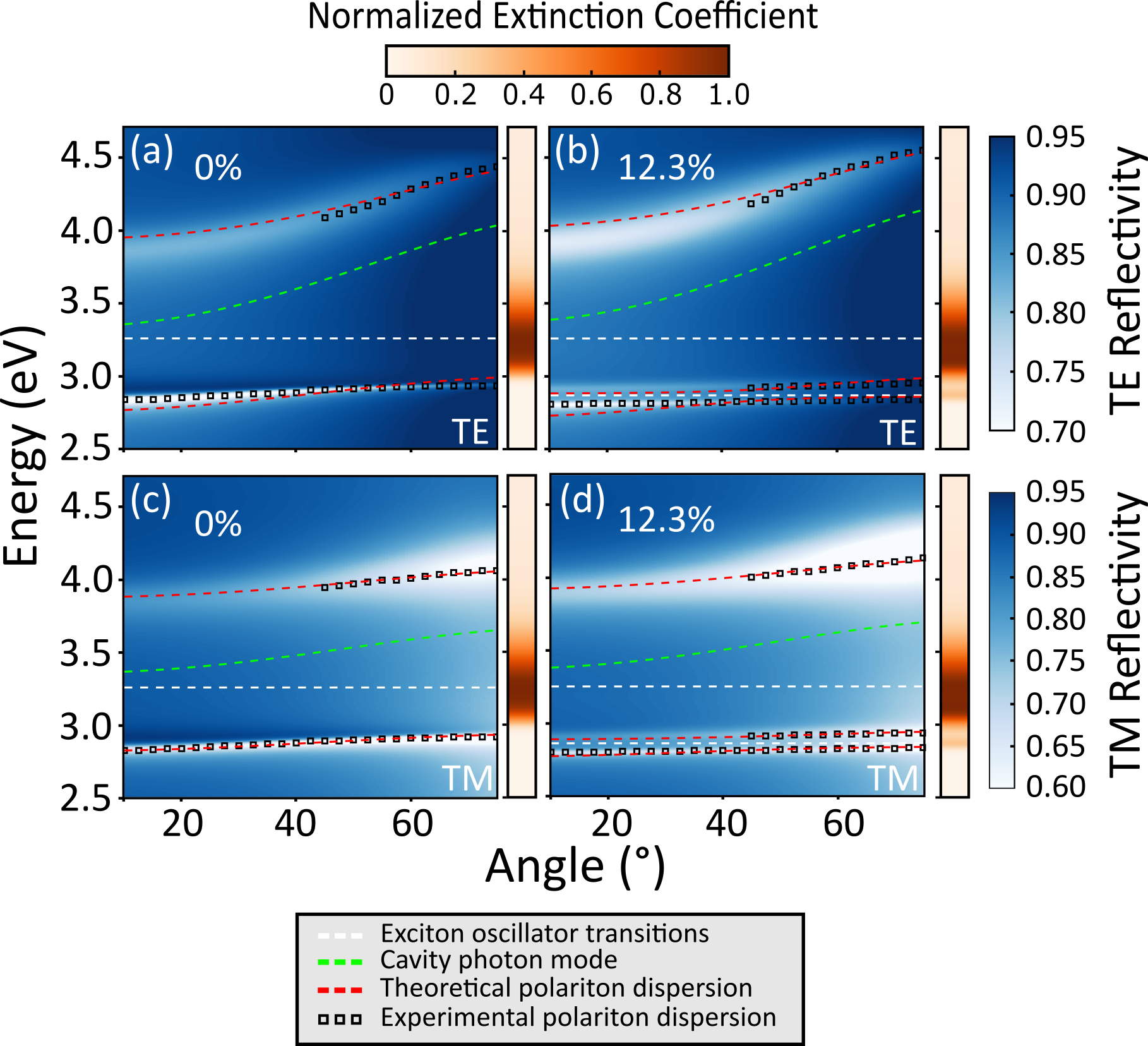}
\end{flushleft}
\caption{\label{fig:Figure5} Experimental and calculated exciton, cavity, and polariton angular dispersions for the 0$\%$ ((a) $\&$ (c)) and 12.3$\%$ ((b) $\&$ (d)) cavities for both TE- ((a) $\&$ (b)) and TM-polarization ((c) $\&$ (d)). The background to each panel comprises the associated R-map obtained from TMR calculations (Section III), on top of which are overlaid exciton oscillator energies (horizontal white dashed lines), experimental (black squares) polariton branch, and calculated coupling-free cavity photon mode (green dashed line) dispersions. The black squares data are LP PL maxima (FIG.~\ref{fig:Figure4}) for incidence angles between 15 and 45$	^{\circ}$ and R-minima (FIG.~\ref{fig:Figure3}) between 45 and 75$^{\circ}$, with $M_q$ polariton eigenvalue fits shown by red dashed lines. The normalized in-plane extinction coefficients (FIG.~\ref{fig:Figure2}) for bare films with corresponding $\beta$-phase fractions are shown to the right of each panel. 
}
\end{figure}

Table 1 collects together pre-set and extracted parameter values for the 0 and $\geq 3.7 \%$ cavities. Note that the fitted values for $n_{eff_{TE}} \sim 1.7$ are close to the background index of glassy-phase PFO (FIG.~\ref{fig:Figure2} (a)), whereas $n_{eff_{TM}} \sim 2.5$ is significantly higher, most likely as a result of the additional coupling of the TM-polarized photon mode to the Metal-Insulator-Metal (MIM) plasmon mode supported by the cavity.\cite{r20, r62, r63} FIG.~\ref{fig:Figure5} further presents (experimental and calculated) exciton, cavity, and polariton angular dispersions for the 0$\%$ ((a) $\&$ (c)) and 12.3$\%$ ((b) $\&$ (d)) cavities for both TE- ((a) $\&$ (b)) and TM-polarization ((c) $\&$ (d)). This shows a satisfactory agreement between the theoretical calculation and experimental measurements.

\newpage
\newpage

\begin{table}
\caption{\label{tab:1}Extracted and pre-set parameter values for microcavities containing various $\beta$-phase fractions, as modelled in Section IV. The values are shown for both TE- and TM-polarization.}
\begin{ruledtabular}
\begin{tabular}{c|c|c|c|c|c|c}
$\beta$-phase content: $\%_{\beta}$ ($\%$) & 0	& 3.7 $\pm$ 0.5 & 10.0 $\pm$ 0.5 &	12.3 $\pm$ 0.5 &	13.9 $\pm$ 0.5 & 15.8 $\pm$ 0.5\\
$\hbar\omega_{2} (eV)$\footnote{Exciton oscillator 2 transition energy.} &	- &	2.87 & 2.87 & 2.86 & 2.86 & 2.86\\
$\hbar\Omega_{01_{TE}} (meV)$\footnote{TE-polarized Rabi energy associated with exciton 1 for $\omega_{cav_{TE}}=\omega_{1}$ (see definition in the text).} & 1180 $\pm$ 20 & 1230 $\pm$ 20 &	1230 $\pm$ 20 & 1270 $\pm$ 20 & 1290 $\pm$ 20 & 1300 $\pm$ 20\\
$\hbar\Omega_{01_{TM}} (meV)$\footnote{TM-polarized Rabi energy associated with exciton 1 for $\omega_{cav_{TM}}=\omega_{1}$}	& 1050 $\pm$ 20 & 1020 $\pm$ 20 & 1100 $\pm$ 20 &	1110 $\pm$ 20 & 1120 $\pm$ 20 &	1140 $\pm$ 20\\
$\hbar\Omega_{02_{TE}} (meV)$\footnote{TE-polarized Rabi energy associated with exciton 2 for $\omega_{cav_{TE}}=\omega_{2}$ (see definition in the text).}	& - &	85 $\pm$ 10	& 126 $\pm$ 15 & 156 $\pm$ 10 & 168 $\pm$ 10 & 172 $\pm$ 10\\
$\hbar\Omega_{02_{TM}} (meV)$ \footnote{TM-polarized Rabi energy associated with exciton 2 for $\omega_{cav_{TM}}=\omega_{2}$ (see definition in the text).}	& - &	100 $\pm$ 10 & 123 $\pm$ 10	& 170 $\pm$ 10 & 175 $\pm$ 10	& 171 $\pm$ 10\\
$ n_{eff_{TE}}$\footnote{Effective refractive index for TE polarization.} & 1.7 $\pm$ 0.1 & 1.6 $\pm$ 0.1 & 1.7 $\pm$ 0.1 & 1.7 $\pm$ 0.1	& 1.7 $\pm$ 0.1	& 1.7 $\pm$ 0.1\\
$ n_{eff_{TM}}$\footnote{Effective refractive index for TM polarization.} & 2.5 $\pm$ 0.1 & 2.5 $\pm$ 0.1 & 2.6 $\pm$ 0.1	& 2.5 $\pm$ 0.1 & 2.4 $\pm$ 0.1	& 2.3 $\pm$ 0.1\\
$ \hbar\omega_{cav_{TE}} (0) (eV)$\footnote{TE-polarized energy of the bare cavity mode at normal incidence.} &	3.33 $\pm$ 0.06	& 3.28 $\pm$ 0.04 & 3.24 $\pm$ 0.06 &	3.37 $\pm$ 0.04	& 3.42 $\pm$ 0.05 & 3.50 $\pm$ 0.1\\ 
$ \hbar\omega_{cav_{TM}} (0) (eV)$\footnote{TM-polarized energy of the bare cavity mode at normal incidence.}	& 3.35 $\pm$ 0.02	& 3.32 $\pm$ 0.04 &	3.30 $\pm$ 0.05 & 3.37 $\pm$ 0.04	& 3.40 $\pm$ 0.05 &	3.42 $\pm$ 0.04\\
\end{tabular}
\end{ruledtabular}
\end{table}

Starting with the 0$\%$ microcavity (FIG.~\ref{fig:Figure5}(a), TE-polarized), the fitted Rabi energy $\hbar\Omega_{01_{TE}}=1.18$ eV is slightly higher but still in good agreement with the value previously reported by Tropf \textit{et al.} (1.09 eV) for a similar PFO cavity.\cite{r57} The coupling ratio $g=\frac{\Omega_{01_{TE}}}{\omega_{1}} =36\%$ clearly exceeds the $\approx 20\%$ value that delineates USC\cite{r20} and thereby justifies use of the full Hopfield Hamiltonian. The value of $\hbar\Omega_{01_{TM}}=1.05$ eV (FIG.~\ref{fig:Figure5}(b), Table I) is somewhat lower than $\hbar\Omega_{01_{TE}}$ due to a strong tendency for PFO chains to lie within the film plane,\cite{r55a,r55b,r56} reducing the out-of-plane oscillator strength (c.f. Section II.B.). The exciton $\left | x_{i,1,q} \right |^{2}$ and photon fractions $\left | w_{i,q} \right |^{2}$ of the LP and UP for TE-polarization are represented by dashed lines in Figure 6(a) and (c). As expected, the LP branch becomes increasingly exciton-like and the UP increasingly photon-like as the angle, i.e. the in-plane wave vector, is increased. Also presented in Figure 6(d) by the dashed line is the content of virtual photons ($\left |y_{LP,q} \right |^{2} + \left | y_{UP,q} \right |^{2}$) in the ground state (GS). This virtual contribution to the GS is negligible in the case of coupling ratios $g <20\%$, leading to a GS approximated well by the vacuum of excitons and photons. When g exceeds 20$\%$, however, a squeezed vacuum can form,\cite{r24} for which the virtual photon content is expected to be preserved even in the presence of high losses.\cite{r61} These photons would, in-principle, be extractable,\cite{r24} although the latter possibility has yet to be demonstrated. 

\begin{figure}[H]
\begin{flushleft}
\includegraphics[scale=0.3]{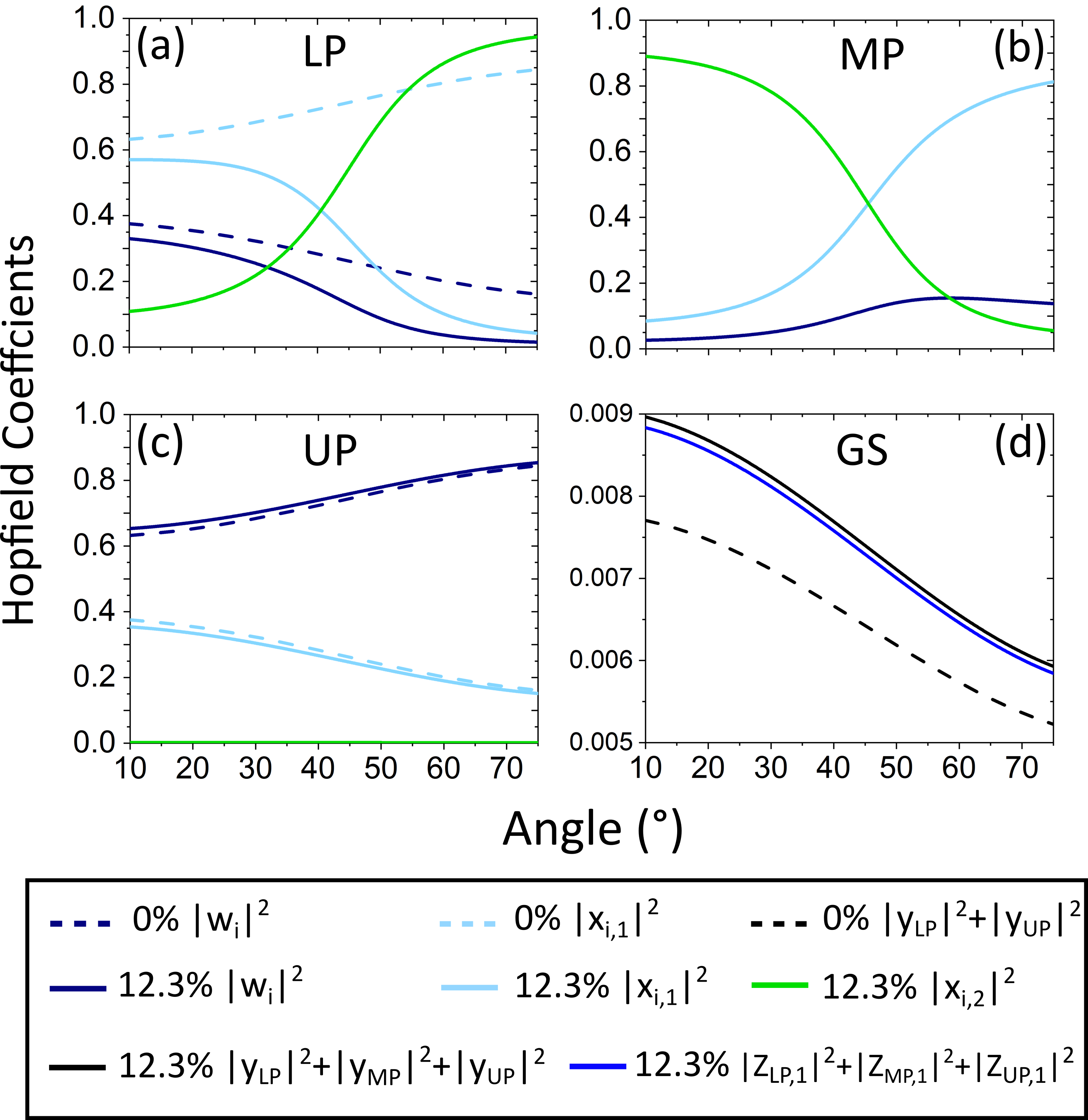}
\end{flushleft}
\caption{\label{fig:Figure6} TE-polarized Hopfield coefficients (see text for definitions) for 0$\%$ and 12.3$\%$ microcavities, obtained using the model presented in Section IV. Results are shown for: (a) LP, (b) MP, (c) UP, and (d) GS. Note the efficient mixing of excitons from populations 1 (solid light blue line) and 2 (solid green line) within the LP (a) and MP (b) of the 12.3$\%$ microcavity.
}
\end{figure}

The 12.3$\%$ cavity (FIG.~\ref{fig:Figure5} (b) and (d)) is representative of the $X_{\beta}\geqslant 3.7\%$ set and shows the already noted (FIG.~\ref{fig:Figure3}(c)) splitting of the 0$\%$ cavity LP into new LP and MP branches. The slight increase (Table I) in the USC $\hbar\Omega_{01_{TE,TM}}$ for this and the other microcavities in the set, is discussed more fully in Section V.B., with specific $X_{\beta}$ contributions considered to more than compensate for the reducing $X_{g}$ absorption coefficient as the two exciton populations interchange. The simultaneous increase (Table 1) in the strong-coupling Rabi energy $\hbar\Omega_{02_{TE}}$ ($\hbar\Omega_{02_{TM}}$) from 85 $\pm$ 10 to 172 $\pm$ 10 meV (100 $\pm$ 10 to 175 $\pm$ 10 meV) on increasing the $X_{\beta}$ fraction from 3.7 to 15.8$\%$ accompanies an evident increase in $X_{\beta}$ extinction coefficient and is also discussed in more detail in Section V.B.. The exciton $\left | x_{i,j,q} \right |^{2}$ and photon $\left | w_{i,q} \right |^{2}$ fractions are presented (solid lines, TE-polarization) in FIG.~\ref{fig:Figure6}(a), (b) and (c) for the LP, MP and UP branches, respectively, of the 12.3$\%$ cavity. A substantial mixing of exciton oscillators 1 and 2 in the LP and MP branches is apparent in FIG.~\ref{fig:Figure6}(a) and (b). In addition, a very small contribution ($\sim 0.2\%$) from exciton oscillator 2 is also present in the UP (FIG.~\ref{fig:Figure6}(c)), highlighting the ability for exciton-polariton physics to create complex energy pathways between polariton levels lying at very different energies. Lastly, as expected from $\hbar\Omega_{02} \ll \hbar\Omega_{01}$, the virtual exciton content in the GS, being the sum of $\left |z_{LP,1,q} \right |^{2} + \left | z_{MP,1,q} \right |^{2} + \left | z_{UP,1,q} \right |^{2}$ for oscillator 1 and $\left |z_{LP,2,q} \right |^{2} + \left | z_{MP,2,q} \right |^{2} + \left | z_{UP,2,q} \right |^{2}$  for oscillator 2, is dominated by the former (blue solid line).

\newpage
\newpage

\section{Discussion}

\subsection{Limitations of the Theoretical Model}

A maximum difference of order 0.1 eV can be observed between the TE-polarization LP fit and experimental data for the 0$\%$ and 12.3$\%$ cavities in FIG.~\ref{fig:Figure5} (a) and (b). Eq.~(\ref{eq:5}), used to determine the photon mode $\omega_{cav_{TE,TM}}(\theta)$ dispersion, is expected to be a good approximation below the light line of the material $\frac{sin^2 (\theta)}{n_{eff}^2} \ll 1)$ and where, as here, the photon energy is far from the metal mirror plasma frequency ($\hbar\omega_{p}>12$ eV for Al).\cite{r20} Note also that if the full trigonometric equations\cite{r62,r63} are used instead, then only minor alterations ( $\approx 0.02$ eV) occur in the computed LP and UP branch dispersions (see Supplemental Material for a full derivation\cite{SI}). This indicates that the use of Eq.~(\ref{eq:5}) is reasonable and that other factors must contribute more to the observed discrepancy. 

Previous studies have noted that the presence of higher lying exciton resonances will affect the polariton dispersion.\cite{r20, r57} Such resonances are clearly visible in Figure 1 (a) and Figure 2 (a) for energies located above 5 eV. TMR calculations using the optical constants presented in FIG.~\ref{fig:Figure2} (a) show that for any photon mode with high enough energy, these resonances yield multiple anti-crossings for which a straightforward analysis is not possible. To test whether adding just a single composite higher energy exciton oscillator, as done in reference 57, can improve the fit to the LP and UP branches (which would then become respectively LP and MP), the eigenvalues of the full matrix $M_{q}$ were fitted to the 0$\%$ microcavity experimental results, with $\hbar\omega_2$ and its Rabi energy $\hbar\Omega_{02}$ allowed to vary respectively in the 4.5 to 6 eV and 0 to 2 eV range. No significant improvement to the fitting residuals was found using this approach, demonstrating that the addition of a single higher energy exciton oscillator does not provide a satisfactory explanation either. 

The spectral shape of the exciton extinction coefficient is expected also to be an important consideration\cite{r64} and this will be discussed in the next section following a more detailed consideration of the Rabi splitting energies $\hbar\Omega_{02_{TE,TM}}$ for the $≥ 3.7\%$ $\beta$-phase microcavities.

\subsection{Rabi Energies and their Dependence on the Oscillator Strengths of the $X_{g}$ and $X_{\beta}$ Populations}

\begin{figure}[H]
\begin{flushleft}
\includegraphics[scale=0.32]{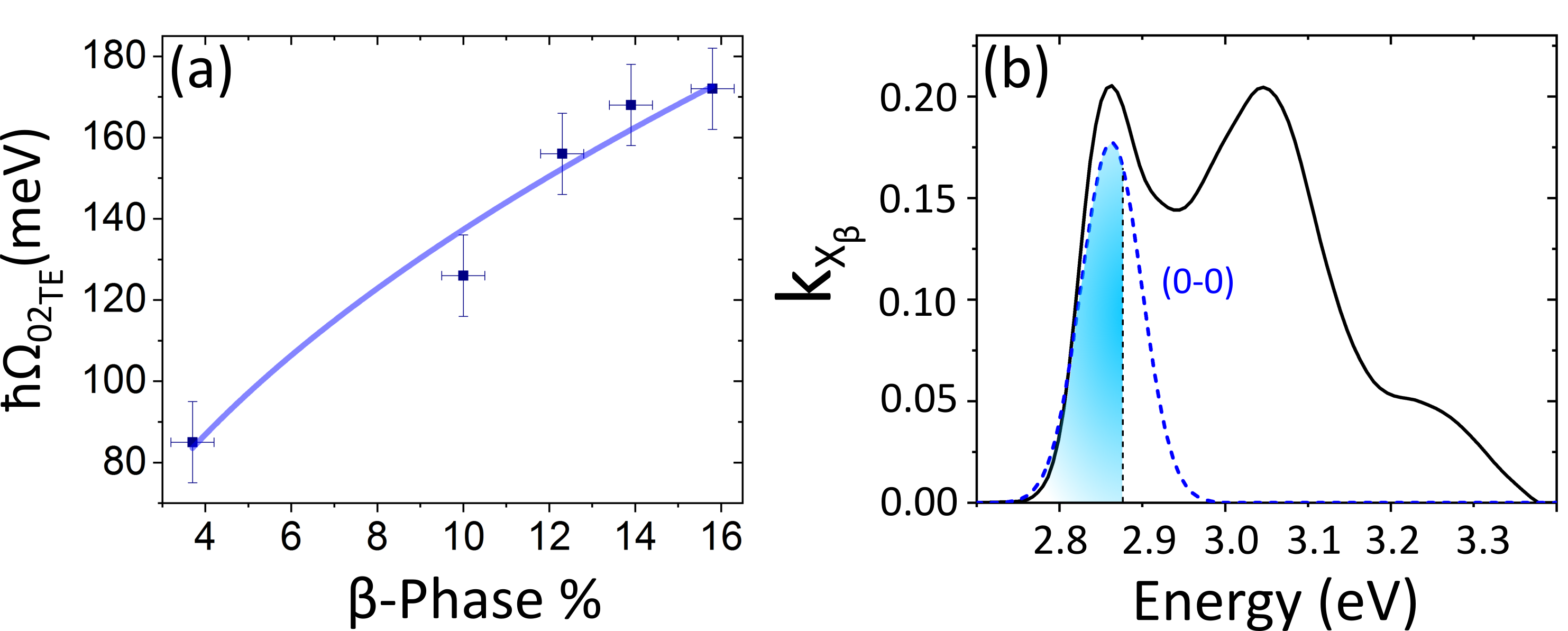}
\end{flushleft}
\caption{\label{fig:Figure7} (a) Rabi splitting energy ($\hbar\Omega_{02_{TE}}$, filled blue squares data) as a function of $\beta$-phase fraction. The solid blue line is the square-root fit ($\hbar\Omega_{02_{TE}}$ =(43.4$\pm$0.9) $\sqrt{\%_{\beta}}$ meV) described in the text. (b) In-plane extinction coefficient, $k_{X_{\beta}}$(E), for the $\beta$-phase component of a 15.8 $\%$ film, extracted from FIG.~\ref{fig:Figure2}(d) following the procedure described in Section V.B. The oscillator strength associated with the light-blue highlighted part of the de-convolved (0-0) vibronic transition (dashed blue line) is sufficient to explain the observed Rabi-splitting ($\hbar\Omega_{02_{TE}}$) between LP and MP. The remainder of the (0-0) together with the (0-1) and (0-2) vibronic transitions can then contribute to the Rabi-splitting ($\hbar\Omega_{01_{TE}}$) between MP and UP, compensating for the interchange in exciton populations between $X_{g}$ and $X_{\beta}$. 
}
\end{figure}

For j separable and spectrally distinct exciton oscillators, the Rabi energies should\cite{r65} scale as
\begin{equation} \label{eq:11}
\hbar\Omega_{0j}\propto\sqrt{N_j f_j},
\end{equation}				                                                            
where $N_j$ and $f_j$ are respectively the number of type-j oscillators and their oscillator strength (see Supplemental Material for a more detailed discussion of this scaling\cite{SI}). For simplicity, it is assumed at first, somewhat na\"ively, that the type-1 and -2 oscillators correspond strictly to $X_{g}$ (j = 1) and $X_{\beta}$ (j = 2) and it is noted that the proportionality constant is identical for both. Next, to allow comparison of the corresponding Rabi energies it is further assumed that the average chromophore lengths for $X_{g}$ and $X_{\beta}$ comprise the same number of monomer units\cite{r53} such that the total number of excitons in the polymer film, $N_0 = N_g + N_{\beta}=\frac{(100-\%_{\beta}) N_0}{100}+\frac{\%_{\beta}N_0}{100}$  where $\%_{\beta}$  is the $\beta$-phase fraction. \ref{eq:11} is then rewritten for both populations:
\begin{equation} \label{eq:12}
\hbar\Omega_{01_{TE,TM}}(100-\%_{\beta})= A_{TE,TM} \sqrt{f_{g_{TE,TM}}}\sqrt{100-\%_{\beta}}
\end{equation}
and
\begin{equation} \label{eq:13}
\hbar\Omega_{02_{TE,TM}}(\%_{\beta})= A_{TE,TM} \sqrt{f_{\beta_{TE,TM}}} \sqrt{\%_{\beta}}, 
\end{equation}   						
with $f_{g_{TE,TM}}$ and $f_{\beta_{TE,TM}}$ the oscillator strengths associated with $X_{g}$ and $X_{\beta}$, respectively, and $A_{TE,TM}$ a population-independent constant. FIG.~\ref{fig:Figure7} (a) shows the resulting fit of Eq.~(\ref{eq:13}) to the $\hbar\Omega_{02_{TE}}$  values from Table 1 for $\%_{\beta} \geqslant 3.7\%$; similar results are obtained for the TM-polarization data. The fit gives:
\begin{equation} \label{eq:14}
\hbar\Omega_{02_{TE}}(\%_{\beta})= (43.4 \pm 0.9) \sqrt{\%_{\beta}} \text{ meV} 
\end{equation}
However, using time-dependent Density Functional Theory (DFT), with the same assumption concerning equal $X_{g}$ and $X_{\beta}$ chromophore length as above, Huang \textit{et al.}\cite{r53} calculated that the ratio $r=\sqrt{\frac{f_{\beta}}{f_g}}$ should be equal to 1.04. For comparison, Eq.~(\ref{eq:12}), ~(\ref{eq:13}),~(\ref{eq:14}) together with Rabi energy $\hbar\Omega_{01_{TE}} (0\%)=1.18$ eV (see Table 1) and the assumption $\sqrt{\frac{f_{\beta}}{f_g}} \approx \sqrt{\frac{f_{\beta_{TE}}}{f_{g_{TE}}}}$, allow derivation of an approximate ‘experimental’ r-value $\approx \frac{\hbar\Omega_{02_{TE}}(100 \%)}{\hbar\Omega_{01_{TE}}(0\%)}=\frac{434}{1180}=0.368$, which is almost three times smaller than the calculated ratio. 

In addition, if $\hbar\Omega_{01_{TE}}$  was indeed only determined by contributions from $X_{g}$, then one would expect this splitting to decrease as the $X_{\beta}$ fraction increased. What is observed, however, is that $\hbar\Omega_{01}$  slightly increases (by $\sim$ 0.1 eV) with increasing $X_{\beta}$ fraction.

Approaching these inconsistencies from a different direction, the experimental Rabi energies and ratio r=0.368 are used to assess in more detail the contributions of $X_{g}$ and $X_{\beta}$ to the exciton-photon coupling. Using the known separability of the $X_{g}$ and $X_\beta$ optical coefficients \cite{r41, r44, r45} a 0.842 (i.e. 1 - $\%_{\beta}$) weighted in-plane $k_{X_g}(E)$ glassy-phase extinction coefficient spectrum (Figure 2 (a)) was subtracted from the corresponding 15.8$\%$ spectrum (Figure 2 (d)) to yield an effective $k_{X_{\beta}}(E)$ spectrum (FIG.~\ref{fig:Figure7} (b)). As expected,\cite{r37a,r37b,r37c} the spectral distribution of $k_{X_{\beta}}(E)$ shows a clearly resolved vibronic progression with $S_{0}-S_{1}$ (0-0), (0-1) and (0-2) vibronic peaks. Among these, the (0-1) and (0-2) vibronic peaks overlap fully with $k_{X_g}(E)$ which falls to zero at around 2.91 eV, with only the (0-0) lying at lower energy. As a consequence, one might anticipate that the $X_{\beta}$ vibronic peaks play different roles. 

The $X_{g}$ and $X_{\beta}$ oscillator strengths are given by:

\begin{equation} \label{eq:15}
f_{g,\beta}\propto\int k_{X_g,X_{\beta}}(E)dE,					        
\end{equation}

where the proportionality constant does not depend on the conformation. Considering that the MP (located at around 2.90 eV in FIG.~\ref{fig:Figure5}(d)) lies below the low energy cut-off for $k_{X_{g}}(E)$ one may reasonably assume that the oscillator strength contributing to the splitting between MP and LP derives entirely from $X_{\beta}$:

\begin{equation} \label{eq:16}
N_{2}=N_{\beta}. 					
\end{equation}		                                     

The oscillator strength $f_{\beta}$ is, however, too large (by a factor of 8) to explain the magnitude of $\hbar\Omega_{02_{TE}}$ obtained from FIG.~\ref{fig:Figure7} (a).  Instead, one can define an energy $E_l$ such that the necessary oscillator strength $f_2$ is the partial integral of $k_{X_{\beta}}(E)$ from 2.7 to $E_l$ eV:

\begin{equation} \label{eq:17}
f_2 = \frac{\int_{2.7}^{E_{l}}k_{X_{\beta}}(E)dE}{\int_{2.7}^{3.4}k_{X_{\beta}}(E)dE}f_{\beta} = R(E_{l})f_{\beta}. 
	                                                                                 \end{equation}

Conversely, as already noted above, the variation of $\hbar\Omega_{01}$  with $X_{\beta}$ fraction tells us that

\begin{equation} \label{eq:18}
N_1= N_g+N_{\beta}=N_0,                                                                                                     
\end{equation}

and that $f_1$ combines a weighted sum of $X_g$ and $X_{\beta}$ contributions, comprising the whole of $f_g$ and a fractional part of $f_{\beta}$, namely $f_{\beta} - f_2$ or equivalently the partial integral of  $k_{X_{\beta}}(E)$ across the range from $E_l$ to 3.4 eV:
\begin{equation} \label{eq:19}
f_1 =\frac{N_g f_g+N_{\beta} (1-R(E_l))f_{\beta}}{N_0} =\begin{bmatrix}\frac{100-\%_{\beta}}{100r^2}+\frac{\%_{\beta}(1-R(E_l))}{100}\end{bmatrix}f_{\beta}.
\end{equation}
From Eq.~(\ref{eq:11}) and Eq.~(\ref{eq:16}) to Eq.~(\ref{eq:19}) we obtain:
\begin{equation} \label{eq:20}
\frac{\hbar\Omega_{02_{TE,TM}}}{\hbar\Omega_{01_{TE,TM}}} =\sqrt{\frac{N_2 f_2}{N_1 f_1}}=\sqrt{\frac{\%_{\beta} f_2}{100 f_1}}=\sqrt{\frac{\%_{\beta} R(E_{l})}{\frac{100-\%_{\beta}}{r^2}+\%_{\beta}(1-R(E_{l}))}}.				              
\end{equation}
Numerically solving Eq.~(\ref{eq:20}) for the 15.8$\%$ cavity with $\hbar\Omega_{01_{TE}}=1.30$ eV and $\hbar\Omega_{02_{TE}}=172$ meV (Table 1) leads to $E_l  \approx 2.85 $ eV. This doesn’t provide any particular intuition other than confirming that a relatively small fraction of the overall $X_{\beta}$ oscillator strength actually contributes to the MP to LP splitting. 
A different and perhaps more physical approach would be to ask: What fraction of the (0-0) vibronic peak is needed? This of course implicitly imposes a dominant role for the (0-0) peak, an action that is supported by the rationale used above to conclude that $N_2=N_{\beta}$, namely that only $k_{X_{\beta}} (E)$ extends in energy below the MP at $\approx 2.90$ eV. Here we add the extension that of $k_{X_{\beta}}(E)$ it is only the (0-0) vibronic that so extends.  The question is then readily addressed by fitting the $k_{X_{\beta}}(E)$ spectrum to Gaussians representing the three resolved vibronic peaks. The resultant (0-0) peak fit (dashed blue line) and the highlighted light blue region representing the required contribution are shown in FIG.~\ref{fig:Figure7}(b). Taking this approach also helps to rationalize the chosen values for $\hbar\omega_{2}$ in Section IV as they are the corresponding (0-0) vibronic peak energies. The remainder of the (0-0) together with the (0-1) and (0-2) vibronic transitions can then contribute to the Rabi-splitting ($\hbar\Omega_{01_{TE}}$) between MP and UP, compensating, as already noted, for the interchange in exciton populations between $X_{g}$ and $X_{\beta}$. 

Whilst this phenomenological approach gives us a reasonable feel for the factors that lie behind the observed behaviour, the precise nature of their contributions remains somewhat ambiguous, mainly because the Agranovich-Hopfield model is not fully adapted to the physics of strongly coupled molecular microcavities. A more suitable theoretical approach with which to study the key parameters and their interplay is needed.  This is also an area in which additional experimental studies can be envisaged, to provide data against which developing theoretical models can then be tested.

Most models used in the spectroscopic analysis of strongly coupled molecular microcavities have either placed resolved vibronic peaks on the same footing as individual exciton transitions by adding an equivalent number of oscillators to the Hamiltonian,\cite{r14,r15,r16,r17,r18} or tried to account for unresolved-vibronic-level-induced absorption asymmetries using broadening terms.\cite{r64,r65} These models, close to that used here, are efficient, well-adapted to fitting procedures and explain to reasonable accuracy the time-integrated reflectivity and photoluminescence spectra. It is clear, however, that they emerge in a rather \textit{ad-hoc} way without a precise knowledge of the underlying molecular physics.

The present study provides a more complex scenario in which the extinction coefficient comprises spectrally overlapping contributions from the $X_{g}$ and $X_{\beta}$ populations with, respectively unresolved and resolved vibronic progressions. In addition, the weighting of the two contributions adjusts as the $\beta$-phase percentage varies, making simple asymmetry corrections problematic. The description of the microscopic features presented above is consequently only partially complete.

In a 2015 report, in order to stimulate new experimental and theoretical investigations, George \textit{et al.}\cite{r49} underlined several observations that remained largely unexplained. For example, they noted that radiative lifetimes integrated over the LP absorbance can be up to three orders of magnitude\cite{r49} longer than those predicted from the LP FWHM (principally determined by the cavity mode FWHM for an inhomogeneously broadened exciton distribution\cite{r59}). In addition, resonant excitation of LP states with 50:50 photon:exciton character often leads to poor LP emission,\cite{r66} suggesting counter-intuitively a very inefficient population of those states. On the other hand, optical pumping of the exciton reservoir, as here, leads to a high PL efficiency.\cite{r67a,r67b} The mechanisms behind such efficient relaxation from the reservoir, proposed to be inelastic exciton-phonon\cite{r68a,r68b,r68c,r68d} or polariton-polariton\cite{r32,r69} (when using high PL efficiency materials including PFO where $QE \leqslant 70\%$\cite{r40,r54}) scattering were not, however, fully understood. Herrera and Spano published a subsequent theory of dark vibronic polaritons\cite{r50a,r50b,r50c} (admixtures of multiple vibronic transitions with single photon states of the cavity) that sought to answer these concerns from a different perspective. Taking into account the roles and interactions of each emitter and accounting for the vibronic couplings existing in the molecular structure through the Holstein term of the Holstein-Tavis-Cummings (HTC) model, they proposed that the emission from the LP (on pumping the exciton reservoir) is enhanced by direct photon leakage from Y-type dark vibronic polariton states located in proximity to the main exciton energy and also close to the upper polariton frequency. 

This may provide a framework with which to more fully explore the role of the two exciton populations for all $\beta$-phase fractions including the 1.1 and 1.5$\%$ microcavities left out of the analysis in Section IV. A further aspect of PFO photophysics to note is the fast non-radiative transfer ($\sim$ps)\cite{r40, r48} between glassy- and $\beta$-phase chain segments that might additionally enhance rapid LP occupation and thereby prevent the simultaneous MP emission seen for other systems.\cite{r32} Applying the HTC model to the dual exciton population ($X_{g}$ plus $X_{\beta}$) PFO system studied here would then require the incorporation of several non-negligible additions that go beyond the purview of the present paper. The reported simulations to date were limited to a single exciton population that was in the strong coupling regime only, already then requiring intensive calculations to represent at most $N_{0} \epsilon [10-20]$.\cite{r50a,r50b,r50c} It will be necessary to extend the current version of the HTC model to include the anti-resonant terms characteristic of USC, then take account of the two different exciton populations, and do so with a relevant number of emitters. In the latter case, in order to provide a good representation for a series of different $\beta$-phase fractions, $N_0 \sim ~ 100$ would probably be necessary. This is not trivial and future work will be needed to implement such additions.

The conformational control approach presented here is of particular interest due to the attractive reservoir material (possessing the same chemical structure as the emitter) and the ability to readily tune optical constants.  The emission characteristics can then be further engineered to refine the already attractive USC effects, yielding, for example, essentially dispersion-free LP emission for both polarizations at modest fractions (e.g. 12.3 $\%$) of $\beta$-phase chain segments. Furthermore, as other studies have shown,\cite{r20, r22, r25a, r25b, r26, r27} the metal-polymer-metal microcavity structure employed also provides a format suited to the electrical injection of carriers, thereby offering the prospect of efficient, angle-insensitive, high colour saturation, wholly-polarized LEDs with emission coming from the dispersion-free LP branch.\cite{r8} 

Finally, since the splitting between the LP and MP is governed by varying the intensity of the resolved $\beta$-phase (0-0) vibronic peak, modifying the energy landscape to study vibronically assisted scattering mechanisms towards efficient exciton-polariton lasing and BEC\cite{r70} may be possible. The study of PFO in a high-Q microcavity might allow quantitative understanding of population mechanisms for the LP ground state in organic condensates. 

\section{Conclusion}

A detailed study of ultrastrong coupling is reported for glassy-phase poly(9,9-dioctylfluorene) excitons ($X_{g}$) within metal-polymer-metal cavities. Control is exerted over the exciton-polariton physics via the generation of increasing fractions of $\beta$-phase chain segments, allowing a systematic study of the dependence of the mode characteristics and resulting light emission properties on the relative $X_{g}$ and $X_{\beta}$ populations. $X_{g}$ ultrastrong coupling shows Rabi splitting energies in excess of 1.05 eV (more than 30$\%$ of the exciton transition energy) for both TE- and TM-polarized light. The splitting of the lower polariton branch, induced by $X_{\beta}$ strong coupling, increases with growing $X_{\beta}$ fraction as expected. However, it is only a fraction of the oscillator strength of the $X_{\beta}$ (0-0) vibronic peak that appears to dominate this splitting, with the remainder and higher vibronics (that overlap spectrally with $X_{g}$ absorption) contributing instead to maintenance and indeed partial enhancement of USC, thereby compensating for the reducing $X_{g}$ population (c.f. $\hbar\Omega_{01}$ values in Table 1).

In all cases, photoluminescence emanates from the lowermost polariton branch, allowing conformational control to be exerted over the emission spectrum and angular dispersion. For a cavity with 12.3$\%$ $\beta$-phase the PL was dispersionless within its $\sim$ 50 meV FWHM linewidth, highlighting the possibility to use such structures for high purity, angle-insensitive LEDs.  An equivalent LED emission spectrum to that measured for PL would yield colour coordinates CIE (x, y) = (0.161, 0.018), with dominant wavelength 447 nm and 99$\%$ saturation. 

Experimental results are discussed in terms of the full Hopfield Hamiltonian generalized to the case of two exciton populations ($X_{g}$ and $X_{\beta}$) simultaneously present within a single semiconductor (PFO). The additional importance of taking account of the molecular characteristics of the semiconductor for an accurate description of its strong coupling behaviour is directly considered, in specific relation to the need to include higher lying exciton states and the role of both resolved and un-resolved vibronic structure.

The observed spectral dispersion of the polariton branches is investigated through a Hopfield-like model. Consideration of the variation in Rabi energies as $X_{g}$ and $X_{\beta}$ populations interchange then leads to a better understanding of how these populations are at play behind the couplings. This first pass at investigating the role of microscopic features has generated the realization that a full understanding of conformational control for exciton-polariton physics will require further significant theoretical input. Models that properly take account of vibronic coupling, such as that developed by Herrera and Spano, \cite{r50a,r50b,r50c} are important in this regard. Achieving such understanding is necessary in order to fully exploit the potential of conformational control, both as a versatile test bed for strong coupling theory and as a path towards novel light sources.

\begin{acknowledgments}

The authors thank Professors Henry Snaith and Moritz Riede for access to fabrication facilities and equipment and Drs Sungho Nam and Sameer Vajjala Kesava for fruitful discussions. They also acknowledge funding from the University of Oxford and from the UK Engineering and Physical Sciences Research Council. F.L.R. further thanks Wolfson College and Dr Simon Harrison for the award of a Wolfson Harrison UK Research Council Physics Scholarship.

\end{acknowledgments}


\begin{thebibliography}{99}

\bibitem{r1}
C.~Weisbuch, M.~Nishioka, A.~Ishikawa and Y.~Arakawa, Phys. Rev. Lett. 1992, 69, 3314.

\bibitem{r2}
D.~Sanvitto and S.~K{\'e}na-Cohen, Nat. Mater 2016, 15, 1061-1073.

\bibitem{r3}
D.G.~Lidzey, D.D.C.~Bradley, M.S.~Skolnick, T.~Virgili, S.~Walker and D.M.~Whittaker, Nature 1998, 395, 53-55, 25692.

\bibitem{r4}
D.G.~Lidzey, D.D.C.~Bradley, T.~Virgili, A.~Armitage and M.S.~Skolnick Phys. Rev. Lett.
1999, 82, 16, 3316-3319.

\bibitem{r5}
D.G.~Lidzey, D.D.C.~Bradley, A.~Armitage, S.~Walker and M.S.~Skolnick, Science 2000, 288, 1620-1623.

\bibitem{r6}
P.~Schouwink, H.V.~Berlepsch, L.~D\"ahne and R.F.~Mahrt, J. Lumin. 2001, 94-95, 821-826.

\bibitem{r7}
P.~Schouwink, H.V.~Berlepsch, L.~D\"ahne and R.F.~Mahrt, Chem. Phys. 2002, 285, 113-120.
 
\bibitem{r8}
N.~Takada, T.~Kamata, D.D.C.~Bradley and Appl. Phys. Lett. 2003, 82, 1812-1814.

\bibitem{r9}
R.F.~Oulton, N.~Takada, J.~Koe, P.N.~Stavrinou and D.D.C.~Bradey, Semicond. Sci. Technol. 2003, 18, S419.

\bibitem{r10}
R.J.~Holmes and S.R.~Forrest, Phys. Rev. Lett. 2004, 93, 186404.

\bibitem{r11}
R.J.~Holmes and S.R.~Forrest, Phys. Rev. B 2005, 71, 235203.

\bibitem{r12a}
R.N.~Marks, J.J.M.~Halls, D.D.C. Bradley, R.H. Friend and A.B. Holmes, J. Phys. Condens. Matter 1994, 6, 1379-1394.~

\bibitem{r12b}
S.~Alvarado, P.~Seidler, D.G.~Lidzey and D.D.C.~Bradley, Phys. Rev. Lett. 1998, 81, 1082-1085.

\bibitem{r13}
S.~K{\'e}na-Cohen, M.~Davanço and S.R.~Forrest, Phys. Rev. Lett 2008, 101, 116401.

\bibitem{r14}
S.~K{\'e}na-Cohen and S.R.~Forrest, Nat. Photonics 2010, 4, 6, 371-375.

\bibitem{r15}
J.D.~Plumhof, T.~St\"oferle, L.~Mai, U.~Scherf and R.F.~Mahrt, Nat. Mater. 2014, 13, 247-252.

\bibitem{r16}
K.S.~Daskalakis, S.A.~Maier, R.~Murray and S.~K{\'e}na-Cohen, Nat. Mater. 2014, 13, 271-278.

\bibitem{r17}
G.~Lerario, A.~Fieramosca, F.~Barachati, D.~Ballarini, K.S.~Daskalakis, L.~Dominici, M.D.~Giorgi, S.A.~Maier, G.~Gigli, S.~K{\'e}na-Cohen and D.~Sanvitto, Nat. Phys. 2017, 13, 837-841.

\bibitem{r18}
N.~Bobrovska, M.~Matuszewski, K.S.~Daskalakis, S.A.~Maier and S.~K{\'e}na-Cohen, ACS Photonics 2018, 5, 111-118.

\bibitem{r19}
P.A.~Hobson, W.L.~Barnes, D.G.~Lidzey, G.A.~Gehring, D.M.~Whittaker, M.S.~Skolnick and S.~Walker, Appl. Phys. Lett. 2002, 81, 3519.

\bibitem{r20}
S.~K{\'e}na-Cohen, S.A.~Maier and D.D.C.~Bradley, Adv. Optical. Mater. 2013, 1, 256, 827-833.

\bibitem{r21}
S.~Gambino.~M., A.~Genco, O.~D.~Stefano, S.~Savasta, S.~Patan\`{e}, D.~Ballarini, F.~Mangione, G.~Lerario, D.~Sanvitto and G.~Gigli, ACS Photonics 2014, 1, 10, 1042-1048.

\bibitem{r22}
M.~Mazzeo, A.~Genco, S.~Gambino, D.~Ballirini, F.~Mangione, O.~Di Stefano, S.~Patan\`{e}, S.~Savasta, D.~Sanvitto and G.~Gigli, Appl. Phys. Lett. 2014, 104, 233303.~

\bibitem{r23a}
V.M.~Agranovich, Opt Spektrosk. 1957, 2, 738.

\bibitem{r23b}
J.J.~Hopfield, Phys. Rev. 1958, 112, 1555.

\bibitem{r24}
C.~Ciuti, G.~Bastard and I.~Carusotto, Phys. Rev. B 2005, 72, 115303.

\bibitem{r25a}
A.~Genco, A.~Ridolfo, S.~Savasta, S.~Patan\`{e}, G.~Gigli and M.~Mazzeo, arXiv:1712.09634 2017.

\bibitem{r25b}
A.~Genco, A.~Ridolfo, S.~Savasta, S.~Patan\`{e}, G.~Gigli and M.~Mazzeo, Adv. Opt. Mater. 2018, doi: 10.1002/adom.201800364.

\bibitem{r26}
F.~Barachati, J.~Simon, Y.A.~Getmanenko, S.~Barlow, S.R.~Marder and S.~K{\'e}na-Cohen, ACS Photonics 2018, 5, 119-125.

\bibitem{r27}
M.~Held, A.~Graf, Y.~Zakharko, P.~Chao, L.~Tropf, M.C.~Gather and J.~Zaumseil, Adv. Opt. Mater 2018, 6, 1700962.

\bibitem{r28}
L.~Garziano, A.~Ridolfo, S.~D.~Liberato and S.~Savasta, ACS Photonics 2017, 4, 2345-2351.

\bibitem{r29}
L.C.~Flatten, D.M.~Coles, Z.~He, D.G.~Lidzey, R.A.~Taylor, J.H.~Warner and J.M.~Smith, Nat. Commun. 2017, 8, 14097.

\bibitem{r30}
G.~Accorsi, S.~Carallo, M.~Mazzeo, A.~Genco, S.~Gambino and G.~Gigli, Chem. Commun. 2014, 50, 1122.

\bibitem{r31}
L.C.~Flatten, S.~Christodoulou, R.K.~Patel, A.~Buccheri, D.M.~Coles, B.P.L.~Reid, R.A.~Taylor, I.~Moreels and J.M.~Smith, Nano Lett. 2016, 16, 11, 7137-7141.

\bibitem{r32}
D.M.~Coles, Q.~Chen, L.C.~Flatten, J.M.~Smith, K.~M\"ullen, A.~Narita and D.G.~Lidzey, Nano Lett. 2017, 17, 5521-5525.

\bibitem{r33}
F.~Scafirimuto, D.~Urbonas, U.~Scherf, R.F.~Mahrt, T.~Stöferle and ACS Photonics 2018, 5, 85-89.

\bibitem{r34}
A.W.~Grice, D.D.C.~Bradley, M.T.~Bernius, M.~Inbasekaran, W.W.~Wu and E.P.~Woo, Appl. Phys. Lett. 1998, 73, 629-631.

\bibitem{r35}
L.L.~Chua, J.~Zaumseil, J.-F.~Chang, E.C.-W.~Ou, P.K.-H.~Ho, H.~Sirringhaus and R.H.~Friend, Nature 2005. 434, 194-199.

\bibitem{r36a}
R.~Xia, G.~Heliotis, Y.~Hou, D.D.C.~Bradley and Org. Electron. 2003, 4, 165-177.

\bibitem{r36b}
G.~Heliotis, R.~Xia, D.D.C.~Bradley, G.A.~Turnbull, I.D.W.~Samuel, P.~Andrew and W.L.~Barnes Appl. Phys. Lett. 2003, 83, 2118-2120.

\bibitem{r36c}
B.~K.~Yap, R.~Xia, M.~Campoy-Quiles, P.N.~Stavrinou, D.D.C.~Bradley and Nat. Mater. 2008, 7, 376-380, 2165.

\bibitem{r37a}
D.D.C.~Bradley, M.~Grell, X.~Long, H.~Mellor, A.W.~Grice, M.~Inbasekara and E.P.~Woo, Proc. SPIE 1997, 3145, 254-259.

\bibitem{r37b}
M.~Grell, D.D.C.~Bradley, G.~Ungar, J.~Hill and K.~Whitehead, Macromolecules 1999, 32, 5810-5817.

\bibitem{r37c}
A.J.~Cadby, P.A.~Lane, H.~Mellor, S.J.~Martin, M.~Grell, C.~Giebeler, D.D.C.~Bradley, M.~Wohlgenannt, C.~An and Z.V.~Vardeny, Phys. Rev. B 2000, 62, 15604-15609.

\bibitem{r38}
M.~Grell, D.D.C.~Bradley, X.~Long, T.~Chamberlain, M.~Inbasekaran, E.P.~Woo and M.~Soliman, Acta Polym. 1998, 49, 439-444.

\bibitem{r39a}
A.L.T.~Khan, P.~Sreearunothai, L.M.~Herz, M.J.~Banach and A.~K\"ohler, Phys. Rev. B 2004, 69, 085201.

\bibitem{r39b}
W.~Chunwaschirasiri, B.~Tanto, D.L.~Huber, M.J.~Winokur and Phys. Rev. Lett. 2005, 94, 107402.

\bibitem{r40}
A.~Perevedentsev, N.~Chander, J-S.~Kim and D.D.C.~Bradley, J. Polym. Sci. Pol. Phys. 2016, 54, 1995-2006.

\bibitem{r41}
P.N.~Stavrinou, G.~Ryu, M.~Campoy-Quiles and D.D.C.~Bradley, J.~Phys. Condens. Matter 2007, 19, 466107.

\bibitem{r42a}
T.~Virgili, D.~Marinotto, G.~Lanzani and D.D.C.~Bradley, Appl. Phys. Lett. 2005, 86, 9.

\bibitem{r42b}
G.~Ryu, R.~Xia and D.D.C.~Bradley, J. Phys. : Condens. Matter 2007, 19, 056205. 

\bibitem{r43}
C.~Rothe, F.~Galbrecht, U.~Scherf and A.P.~Monkman, Adv. Mater. 2006, 18, 2137–2140.

\bibitem{r44}
G.~Ryu, P.N.~Stavrinou and D.D.C.~Bradley, Adv. Funct. Mater. 2009, 19, 3237-3242.

\bibitem{r45}
A.~Perevedentsev, Y.~Sonnefraud, S.~Sharma, A.E.G.~Cass, S.A.~Maier, J.S.~Kim, P.N.~Stavrinou and D.D.C.~Bradley, Nat. Commun. 2015, 6, 5977.

\bibitem{r46}
J.~Peet, E.~Brocker, Y.~Xu and G.C.~Bazan, Adv. Mater. 2008, 20, 10.

\bibitem{r47}
Q.~Zhang, L.~Chi, G.~Hai, Y.~Fang, X.~Li, R.~Xia, W.~Huang and E.~Gu, Molecules 2017, 22(2), 315.

\bibitem{r48}

M.~Ariu, M.~Sims, M.D.~Rahn, J.~Hill, A.M.~Fox, D.G.~Lidzey, M.~Oda, J.~Cabanillas-Gonzalez and D.D.C.~Bradley, Phys. Rev. B 2003, 67, 195333.

\bibitem{r49}

J.~George, S.~Wang, T.~Chervy, A.~Canaguier-Durand, G.~Schaeffer, J.M.~Lehn, J.A.~Hutchison, C.~Genet and T.W.~Ebbesen, Farad. Disc. 2015, 178, 281.

\bibitem{r50a}
F.~Herrera and F.C.~Spano, Phys. Rev. A 2017, 95, 053867.

\bibitem{r50b}
F.~Herrera and F.C.~Spano, Phys. Rev. Lett. 2017, 118, 223601.

\bibitem{r50c}
F.~Herrera and F.C.~Spano, ACS Photonics 2018, 5, 65-79.

\bibitem{r51}
M.~Ahsan Zeb, P.G.~Kirton and J.~Keeling, ACS Photonics 2018, 5, 249-257.

\bibitem{r52}
http://www.1-material.com/ 

\bibitem{r53}
L.~Huang, X.~Huang, G.~Sun, C.~Gu, D.~Lu and Y.~Ma, J. Phys. Chem. C 2012, 116, 7993-7999.~

\bibitem{SI}
See Supplemental Material at [URL will be inserted by publisher] for remaining thin films absorption, photoluminescence and optical constants spectra, for remaining microcavity TE/TM-polarized reflectivity maps, for the microcavity photoluminescence spectral characteristics and theoretical considerations on the photon mode and the Rabi energy.

\bibitem{r54}
M.~Ariu, D.G.~Lidzey, M.~Sims, A.J.~Cadby, P.A.~Lane and D.D.C.~Bradley, J. Phys. Condens. Matter 2002, 14, 9975-9986.

\bibitem{r55a}
M.~Campoy-Quiles, G.~Heliotis, R.~Xia, M.~Ariu, M.~Pintani, P.G.~Etchegoin and D.D.C.~ Bradley, Adv. Funct. Mater. 2005, 15, 925-933.

\bibitem{r55b}
M.~Campoy-Quiles, M.~Isabel Alonzo, D.D.C.~Bradley and L.J.~Richter, Adv. Funct. Mater. 2014, 24, 2116-2134.

\bibitem{r56}
M.~Campoy-Quiles, P.G.~Etchegoin and D.D.C.~Bradley, Phys. Rev. B 2005, 72, 045209.

\bibitem{r57}
L.~Tropf, C.P.~Dietrich, S.~Herbst, A.L.~Kanibolotsky, P.J.~Skabara, F.~W\"urther, I.D.~W.~Samuel, M.C.~Gather and S.~H\"ofling, Appl. Phys. Lett. 2017, 110, 153302.

\bibitem{r58}
G.~Panzarini, L.C.~Andreani, A.~Armitage, D.~Baxter, M.S.~Skolnick, V.N.~Astratov, J.S.~Roberts, A.V.~Kakovin, M.R.~Vladimirova and M.A.~Kaliteevski, Phys. Rev. B 1999, 59, 5082-5089.

\bibitem{r59}
R.~Houdr{\'e}, R.P.~Stanley and M.~Ilegems, Phys.~Rev.~A 1996, 53, 2711.

\bibitem{r60a}
T.A.~Fisher, D.G.~Lidzey, M.A.~Pate, M.S.~Weaver, D.M.~Whittaker, M.S.~Skolnick and D.D.C.~Bradley, Appl. Phys. Lett. 1995, 67, 1355-1357.

\bibitem{r60b}
D.G.~Lidzey, D.D.C.~Bradley, S.J.~Martin and M.A.~Pate, IEEE J. Sel. Top. Quantum Electron. 1998, 4, 113-118. 

\bibitem{r61}
S.D.~Liberato, Nat. Commun. 2017, 8, 1465.

\bibitem{r62}
E.N.~Economou, Phys. Rev. 1969, 182 , 539.

\bibitem{r63}
M.~Litinskaya and V.M.~Agranovich, J. Phys. Condens. Matter 2012, 24, 015302.

\bibitem{r64}
A.~Armitage, D.G.~Lidzey, D.D.C.~Bradley, T.~Virgili, M.S.~Skolnick and S.~Walker, Synth. Met. 2000, 111-112, 377-379.

\bibitem{r65}
M.~Fox Quantum Optics - An Introduction (Oxford Master Series) Oxford University Press 2006.

\bibitem{r66}
D.M.~Coles, R.T.~Grant, D.G.~Lidzey, C.~Clark and P.G.~Lagoudakis, Phys. Rev. B: Condens. Matter Mater. Phys. 2013 88, 121303.

\bibitem{r67a}
P.~Michetti and G.C.~La Rocca, Phys. Rev. B: Condens. Matter Mater. Phys. 2009, 79, 035325.

\bibitem{r67b}
M.~Litinskaya, P.~Reineker and V.M.~Agranovich, J. Lumin. 2004, 110, 364-372.

\bibitem{r68a}
D.M.~Coles, P.~Michetti, C.~Clark, W.C.~Tsoi, A.M.~Adawi, J.-S.~Kim and D.G.~Lidzey, Adv. Opt. Mater. 2011, 21, 3691-3696.

\bibitem{r68b}
M.~Litinskaya, P.~Reineker and V.M.~Agranovich, J. Lumin. 2006, 119-120, 277-282.

\bibitem{r68c}
P.~Michetti and G.C.~La Rocca, Phys. Rev. B: Condens. Matter Mater. Phys. 2008, 77, 195301.

\bibitem{r68d}
N.~Somaschi, L.~Mouchliadis, D.M.~Coles, I.~Perakis, D.G.~Lidzey, P.~Lagoudakis and P.~Savvidis, Appl. Phys. Lett. 2011, 99, 143303.

\bibitem{r69}
R.T.~Grant, P.~Michetti, A.J.~Musser, P.~Gregoire, T.~Virgili, E.~Vella, M.~Cavazzini, K.~Georgiou, F.~Galeotti, C.~Clark, J.~Clark, C.~Silva and D.G.~Lidzey, Adv. Opt. Mater.~2016, 4, 1615−1623.

\bibitem{r70}
L.~Mazza, S.~K{\'e}na-Cohen, P.~Michetti and G.C.~La Rocca, Phys. Rev. B 2013,  88, 075321.

\end{thebibliography}
\end{document}